\documentclass[fleqn,usenatbib]{mnras}

\usepackage{newtxtext,newtxmath}

\usepackage[T1]{fontenc}

\DeclareRobustCommand{\VAN}[3]{#2}
\let\VANthebibliography\thebibliography
\def\thebibliography{\DeclareRobustCommand{\VAN}[3]{##3}\VANthebibliography}

\usepackage{graphicx}	
\usepackage{amsmath}	
\usepackage{amssymb}	

\title[DECam DDF 2022]{Deep Drilling in the Time Domain with DECam: Survey Characterization}

\author[M. L. Graham al.]{Melissa~L.~Graham$^{1}$\thanks{E-mail: mlg3k@uw.edu}, 
Robert A. Knop$^{2}$,
Thomas Kennedy$^{1,3}$, 
Peter E. Nugent$^{2}$,
Eric Bellm$^{1}$, 
\newauthor
M{\'a}rcio Catelan$^{4,5}$, 
Avi Patel$^{6}$,
Hayden Smotherman$^{1}$, 
Monika Soraisam$^{7}$, 
Steven Stetzler$^{1}$, 
\newauthor
Lauren N. Aldoroty$^{8}$,
Autumn Awbrey$^{9}$, 
Karina Baeza-Villagra$^{4,5}$, 
Pedro H. Bernardinelli$^{1}$, 
Federica Bianco$^{10}$,
\newauthor
Dillon Brout$^{11}$, 
Riley Clarke$^{10}$, 
William I. Clarkson$^{12}$, 
Thomas Collett$^{13}$, 
James R. A. Davenport$^{1}$, 
\newauthor
Shenming Fu$^{7}$, 
John E. Gizis$^{10}$, 
Ari Heinze$^{1}$,
Lei Hu$^{14}$, 
Saurabh W. Jha$^{15}$, 
Mario Juri{\'c}$^{1}$, 
J. Bryce Kalmbach$^{1}$, 
\newauthor
Alex Kim$^{2}$, 
Chien-Hsiu Lee$^{16}$, 
Chris Lidman$^{17,18}$, 
Mark Magee$^{13}$, 
Clara E. Mart{\'i}nez-V{\'a}zquez$^{19}$, 
\newauthor
Thomas Matheson$^{7}$, 
Gautham Narayan$^{20}$ 
Antonella Palmese$^{21,22}$,
Christopher A. Phillips$^{1}$,
Markus Rabus$^{23}$,
\newauthor
Armin Rest$^{24,25}$,
Nicol{\'a}s Rodr{\'i}guez-Segovia$^{26}$, 
Rachel Street$^{27}$,
A. Katherina Vivas$^{28}$, 
Lifan Wang$^{8}$, 
\newauthor
Nicholas Wolf$^{7}$, 
and Jiawen Yang$^{8}$ 
\\
$^{1}$DIRAC Institute, Department of Astronomy, University of Washington, 3910 15th Avenue NE, Seattle, WA 98195, USA\\
$^{2}$E.O. Lawrence Berkeley National Laboratory, 1 Cyclotron Rd., Berkeley, CA, 94720 \\
$^{3}$Department of Astronomy, University of Michigan, 1085 S. University, 323 West Hall, Ann Arbor, MI 48109-1107, USA \\
$^{4}$Pontificia Universidad Cat{\'o}lica de Chile, Av. Vicu{\~n}a Mackenna 4860, 7820436 Macul, Santiago, Chile \\
$^{5}$Millennium Institute of Astrophysics, Nuncio Monse{\~n}or S{\'o}tero Sanz 100, Providencia, Santiago, Chile \\  
$^{6}$Department of Physics and Astronomy, Haverford College, Haverford, PA 19041, USA \\
$^{7}$NSF's National Optical-Infrared Astronomy Research Laboratory, 950 N. Cherry Avenue, Tucson, AZ, 85719, USA\\
$^{8}$George P. and Cynthia Woods Mitchell Institute for Fundamental Physics and Astronomy, Department of Physics and Astronomy, Texas A\&M University, College Station, TX 77843, USA \\
$^{9}$Department of Astronomy, University of California, Berkeley, CA 94720-3411, USA \\
$^{10}$Department of Physics and Astronomy, University of Delaware, Newark, DE 19716, USA \\
$^{11}$Center for Astrophysics | Harvard \& Smithsonian, 60 Garden Street, Cambridge, MA 02138, USA \\
$^{12}$Department of Natural Sciences, University of Michigan-Dearborn, 4901 Evergreen Road, Dearborn, MI 48128, USA \\
$^{13}$Institute of Cosmology and Gravitation, University of Portsmouth, Burnaby Rd, Portsmouth PO1 3FX, UK \\
$^{14}$Purple Mountain Observatory, Nanjing 210023, People’s Republic of China \\
$^{15}$Department of Physics and Astronomy, Rutgers the State University of New Jersey, 136 Frelinghuysen Road, Piscataway, NJ 08854, USA\\
$^{16}$W. M. Keck Observatory, Kamuela, HI 96743, USA\\
$^{17}$The Research School of Astronomy and Astrophysics, The Australian National University, ACT 2601, Australia \\
$^{18}$Centre for Gravitational Astrophysics, College of Science, The Australian National University, ACT 2601, Australia \\
$^{19}$ Gemini Observatory, NSF’s NOIRLab, 670 N. A’ohoku Place, Hilo, HI 96720, USA \\
$^{20}$Department of Astronomy, University of Illinois at Urbana-Champaign, 1002 W. Green St, IL 61801, USA \\
$^{21}$Department of Physics, University of California Berkeley, 366 LeConte Hall MC 7300, Berkeley, CA, 94720, USA\\ 
$^{22}$NASA Einstein Fellow\\
$^{23}$Departamento de Matem{\'a}tica y F{\'i}sica Aplicadas, Facultad de Ingenier{\'i}a, Universidad Cat{\'o}lica de la Sant{\'i}sima Concepci{\'o}n, Alonso de Rivera 2850, Concepci{\'o}n, Chile \\
$^{24}$Space Telescope Science Institute, 3700 San Martin Dr., Baltimore, MD 21218, USA \\
$^{25}$Physics and Astronomy Department, Johns Hopkins University, Baltimore, MD 21218, USA \\
$^{26}$School of Science, University of New South Wales, Australian Defence Force Academy, Canberra, ACT 2600, Australia \\
$^{27}$Las Cumbres Observatory, 6740 Cortona Drive, Suite 102, Goleta, CA 93117, USA \\
$^{28}$Cerro Tololo Inter-American Observatory/NSF’s NOIRLab, Casilla 603, La Serena, Chile 
}

\date{Accepted XXX. Received YYY; in original form ZZZ}

\pubyear{2022}

\begin{document}
\label{firstpage}
\pagerange{\pageref{firstpage}--\pageref{lastpage}}
\maketitle

\begin{abstract}
This paper presents a new optical imaging survey of four deep drilling fields (DDFs), two Galactic and two extragalactic, with the Dark Energy Camera (DECam) on the 4 meter Blanco telescope at the Cerro Tololo Inter-American Observatory (CTIO).
During the first year of observations in 2021, $>$4000 images covering 21 square degrees (7 DECam pointings), with $\sim$40 epochs (nights) per field and 5 to 6 images per night per filter in $g$, $r$, $i$, and/or $z$, have become publicly available (the proprietary period for this program is waived).
We describe the real-time difference-image pipeline and how alerts are distributed to brokers via the same distribution system as the Zwicky Transient Facility (ZTF).
In this paper, we focus on the two extragalactic deep fields (COSMOS and ELAIS-S1), characterizing the detected sources and demonstrating that the survey design is effective for probing the discovery space of faint and fast variable and transient sources.
We describe and make publicly available 4413 calibrated light curves based on difference-image detection photometry of transients and variables in the extragalactic fields.
We also present preliminary scientific analysis regarding Solar System small bodies, stellar flares and variables, Galactic anomaly detection, fast-rising transients and variables, supernovae, and active galactic nuclei.
\end{abstract}


\section{Introduction}

When Rubin Observatory begins operations in a few years, the time-domain data on transients and variables will come from both the Legacy Survey of Space and Time's (LSST’s) wide-fast-deep (WFD) main survey and its deep drilling fields (DDF), providing a rich ecosystem of detections at different depths and timescales \citep{2019ApJ...873..111I}.
The current leading precursor survey for the LSST WFD is the Zwicky Transient Facility (ZTF; \citealt{2019PASP..131g8001G,2019PASP..131a8002B,2019PASP..131a8003M}): its Northern Sky Survey covers nearly the entire visible sky every second night to $\sim$20.5 mag in the $g$- and $r$-filters \citep{Bellm:19:ZTFScheduler}.
The ZTF’s real-time difference imaging and analysis pipeline produces a public alert stream based on the same alert packet format and distribution mechanism as developed for the LSST \citep{2019PASP..131a8001P}.

In order to take another step towards the Rubin era, and enrich our current time-domain alert ecosystem, we are conducting an imaging survey of four DDFs with the Dark Energy Camera (DECam; \citealt{2008SPIE.7014E..0ED}), with real-time data processing and public alert distribution using the same system as the ZTF. 
The proprietary period for the images from this DECam DDF program has been and will continue to be waived, to allow public access.
As we describe in this paper, this DDF survey produces a sequence of alerts in multiple filters in four small-area regions of sky, presenting a new challenge for alert brokers and time-domain astronomers, and the opportunity to identify fast-changing transients and variables during the night.
The main science goals of this survey are to obtain a better understanding of faint and fast variable and transient sources (e.g., supernovae, GRB afterglows, Galactic novae, microlensing events, flares) by generating well-sampled multi-band light-curves.

In Section~\ref{sec:survey} we present the survey design and the selected fields, and characterize the observing strategy performance in terms of inter-night gaps and image quality.
In Section~\ref{sec:proc} we describe the image processing and alert generation system, and characterize the resulting image quality and source detection capabilities (e.g., the real/bogus score).
A set of 4413 high-quality candidates (time-series of difference-image detections associated by coordinate; i.e., light curves) is presented and made publicly available for scientific analysis.
In Section \ref{sec:sci}, we provide a few examples of preliminary, ongoing scientific investigations with the DDF data.

\section{Survey Design}\label{sec:survey}

In the original proposal for "Deep Drilling in the Time Domain with DECam", we attempted to fit into the half-night classical scheduling of CTIO by designing a strategy in which an extragalactic and a Galactic field would be observed every other night (i.e., a two-day cadence) in 20 to 30 minute windows immediately after/before evening/morning twilight, or near midnight.

However, this original plan was not feasible, and instead this DDF program became one of the foundational programs for the DECam Alliance for Transients (DECAT), a group of DECam principal investigators (PIs) with time-domain programs, all of which only require up to a couple of hours per night, who request to be co-scheduled for shared full or half nights.
The PIs work together to create observing plans that include targets from all programs, thereby enabling dynamic queue-like scheduling for an otherwise classically scheduled facility.
The DECam observation scripts include the individual proposal identifiers so that each PI can track their own time usage, apply their own proprietary period, and process their own data.
Together, the DECAT programs were co-scheduled for every $\sim$3rd night from March 18 through June 10 in 2021A, and September 16 through January 23 in 2021B.

The proprietary period was waived for images obtained as part of the DECam DDF program (this is not the case for all of the other programs being co-scheduled under DECAT).
This work uses only images from the DECam DDF program obtained in 2021, all of which are available in the NOIRLab archive by searching for proposal identifier 2021A-0113 and 2021B-0149.
DECam DDF images obtained in 2022 are available under proposal identifiers 2022A-724693 and 2022B-762878.

\subsection{Field Selection and Exposure Times}\label{ssec:survey_fieldsel}

\subsubsection{Two Extragalactic DDFs: COSMOS and ELAIS}\label{sssec:extgalddfs} 

The COSMOS field was chosen as one of the two DDF extragalactic fields due to its legacy value \citep{2007ApJS..172....1S}: COSMOS has been observed by many programs in the past and has been selected as one of the future LSST DDFs.
Because another DECAT program also observes COSMOS for their own distinct science goals, in order to maximize the scientific utility of the DECam DDF images we slightly modified our COSMOS fields from the original proposal in order to use the exact same three pointings (field center coordinates), as listed in Table~\ref{tab:1} as COSMOS-1, -2, and -3 (and which we collectively refer to as COSMOS hereafter).

Every night that the DDF program observed COSMOS, three $\sim$0.5 hour (non-identical) sequences were done, separated in time when possible but often done back-to-back (see \S~\ref{ssec:survey_char}).
The three sequences, A, B, and C, cycle over the fields obtaining a series of images in the $g$, $r$, and $i$ filters with exposure times of 60, 86, and 130 seconds, respectively.
Sequence A does COSMOS-1 in $gri$, then COSMOS-2 in $gri$, then COSMOS-3 in $gri$, then COSMOS-1 in $gri$, then COSMOS-2 in $gri$; Sequence B starts with COSMOS-3 and then does fields 1, 2, 3, and 1, always doing all three filters after every slew; and then finally Sequence C starts with COSMOS-2 and does fields 3, 1, 2, and 3.
In total, each field was imaged five times in each filter, for a total of 15 photometric observations per night (45 for sources in the $\sim$5\% overlap region; no dithering).
The 5-sigma limiting magnitudes are r$\sim$23.5 mag (single exposure) and r$\sim$24.5 mag (nightly stack).
Similar limits are achieved for the $g$ and $i$ filters, too, which was the main motivation for the adopted exposure times.

As the COSMOS field began to set in 2021A (late May) we adjusted the strategy to obtain just one image per filter per field per night, with exposure times that matched a long-term AGN monitoring program for COSMOS (80, 70, and 90 seconds in the $g$, $r$, and $i$ filters, respectively).
This strategy allowed the AGN science to continue and for us to obtain a few more epochs (nights in which we visit COSMOS) for the DDF program.

The choice of observing the COSMOS field also has the benefit of being targeted by the Dark Energy Spectroscopic Instrument (DESI; \citealt{desiI,desiII}).
The resulting joint DECam-DESI observations using images from both the DDF described in this work, and the DECam Survey of Intermediate Redshift Transients (DESIRT, \citealt{2022TNSAN.107....1P}; another DECAT program) data will be described in Palmese et al., in preparation.

The second extragalactic field is the ELAIS-S1\footnote{ELAIS: European Large Area Infrared Space Observatory (ISO) Survey; \citep{2000MNRAS.316..749O}.} deep field.
We chose two field pointings, which are listed in Table~\ref{tab:1} as ELAIS-E1 and ELAIS-E2 (and which we collectively refer to as ELAIS hereafter).
These pointings match the same central coordinates of another DECAT program, and match two of the ten fields monitored by the Dark Energy Survey's Supernova program (DES-SN; \citealt{2020AJ....160..267S}) -- so these fields have a history of observations since 2013.
Observations of the ELAIS DDF began in late May 2021.
The exposure times for the ELAIS sequences were the same as the COSMOS field: two $\sim$0.5 hour sequences alternated between the fields, obtaining a series of images in the $g$, $r$, and $i$ filters with exposure times of 60, 86, and 130 seconds, respectively.

\subsubsection{Two Galactic DDFs in DECaPS}\label{sssec:galddfs}

During 2021A, we observed a single pointing in the Galactic bulge, within the Dark Energy Camera Plane Survey (DECaPS) region, which we called “DECaPS East” (Table~\ref{tab:1}; galactic coordinates $l=1.462\ b=-3.681$).
This field was chosen to \textit{half}-overlap with field B1 from \citet{2019ApJ...874...30S}, to provide some legacy value and also some new variables.
For 2021B we added another single pointing in DECaPS called “DECaPS West” (Table~\ref{tab:1}; galactic coordinates $l=242.2\ b=-0.91$).
This field was chosen to optimize the creation of template images, as it coincided well with existing DECaPS coverage in the region.

Every night that the DECAT programs were co-scheduled, two $\sim$0.3 hour sequences were done of one or both DECaPS DDFs, usually back-to-back but separated in time on occasion. 
Each sequence cycled through the $g$, $r$, $i$, and $z$ filters three times with exposure times of 96, 50, 30, and 30 seconds, respectively.
In total, each field was imaged six times in each filter for a total of 24 photometric observations in a given night.
The 5-sigma limiting magnitudes are, at most, r$\sim$23.5 mag (single exposure) and r$\sim$24.5 mag (nightly stack) -- but potentially shallower in some regions of these crowded fields.

The DECAT nights were scheduled near full moon, and on some of those nights we attempted to mitigate the high sky background by doing more short exposures with the $g$, $r$, $i$, and $z$ filters: 10, 25, 30, and 30 seconds, respectively.
This was done so rarely that results from these short exposure time images are not further discussed in this paper, but the impact that observing near full moon on the source detection rate is discussed in \S~\ref{sssec:survey_char_bright}.

\begin{table}
\centering
\caption{Field Centers.}
\label{tab:1}
\begin{tabular}{lll}
\hline
Field & RA & Dec \\
 & [$h:m:s$] & [$^{\circ}:\arcmin:\arcsec$] \\
\hline
COSMOS-1 & 10:00:00 & +03:06:00 \\
COSMOS-2 & 09:56:53 & +01:45:00 \\
COSMOS-3 & 10:03:7 & +01:45:00 \\
ELAIS-E1 & 00:31:30 & -43:00:35 \\
ELAIS-E2 & 00:38:00 & -43:59:53 \\
DECaPS-East & 18:03:34 & -29:32:02 \\
DECaPS-West & 07:45:16.8 & -26:15:00 \\
\hline
\end{tabular}
\end{table}

\subsection{Survey Characterization}\label{ssec:survey_char}

\begin{figure}
\includegraphics[width=\columnwidth]{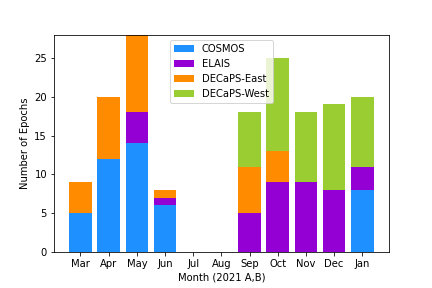}
\caption{The number of epochs per DDF per month, from March 2021 through January 2022 (stacked; as shown in legend). \label{fig:1} }
\end{figure}

In Table \ref{tab:2} we list the number of images obtained by our program, by field and by filter.
We also list the number of epochs, where an epoch is a night in which an image in any filter was obtained.
Figure~\ref{fig:1} provides a bar chart of the number of epochs per field per month that our program obtained over the course of 2021 A \& B semesters.
This figure shows the seasons in which each DDF was observed; these seasons of visibility are constrained by our program's airmass limits.

\begin{table}
\centering
\caption{Number of images by field and filter.}
\label{tab:2}
\begin{tabular}{lcccccc}
\hline
Field & \multicolumn{5}{c}{--------- Number of Images ---------} & Number of \\ 
 & $g$ & $r$ & $i$ & $z$ & Total & Epochs \\ 
\hline
Field & g & r & i & z & Total & Epochs \\ 
COSMOS-1 & 135 & 139 & 143 & 0 & 417 & 41  \\ 
COSMOS-2 & 136 & 139 & 140 & 0 & 415 & 42  \\ 
COSMOS-3 & 132 & 134 & 137 & 0 & 403 & 41  \\ 
ELAIS-E1 & 187 & 186 & 186 & 0 & 559 & 39  \\ 
ELAIS-E2 & 176 & 175 & 173 & 0 & 524 & 39  \\ 
DECaPS-East & 239 & 249 & 257 & 271 & 1016 & 32  \\ 
DECaPS-West & 264 & 261 & 255 & 253 & 1033 & 48  \\ 
\hline
All & 1269 & 1283 & 1291 & 524 & 4367 & 92  \\ 
\hline
\end{tabular}
\end{table}

\begin{figure}
\includegraphics[width=\columnwidth]{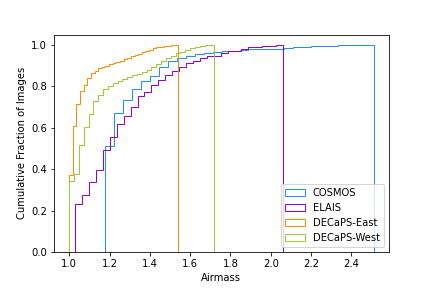}
\caption{The normalized cumulative airmass distributions over all images obtained for our program, for the four fields: COSMOS (blue), ELAIS (purple), DECaPS-East (orange), and DECaPS-West (green).
\label{fig:2} }
\end{figure}

We aimed to observe the fields when they were at airmass $<1.5$ because low airmass observations are better for difference imaging.
The airmass distributions for all of our program's images (Figure~\ref{fig:2}) shows that this goal was achieved for $\gtrsim$80\% of the images for any given field -- even for COSMOS, which is an equatorial field and which reaches a minimum airmass of $\sim1.2$ from CTIO.
The inclusion of high-airmass images for the extragalactic fields is due to our attempts to extend the observing seasons, and to make use of time allocated primarily in second-half nights during 2021B.
As we will discuss in Section~\ref{ssec:sci_gal_flares}, having images at relatively high airmass can lead to additional scientific discoveries.

\begin{figure}
\includegraphics[width=\columnwidth]{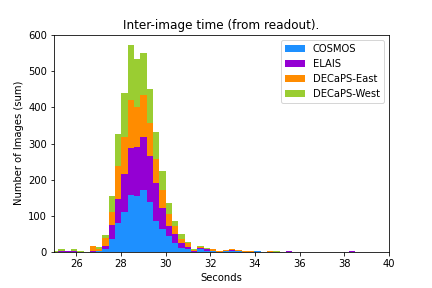}
\includegraphics[width=\columnwidth]{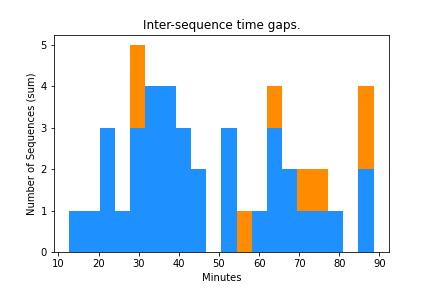}
\includegraphics[width=\columnwidth]{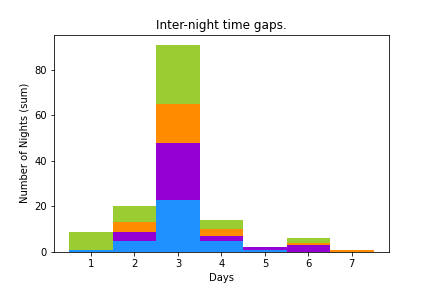}
\caption{Distributions of the times between images. \textit{Top:} the time between successive images (in seconds), which is dominated by readout. \textit{Middle:} the time between DDF sequences within a night, in minutes. \textit{Bottom:} the time between observing nights, in days. Shown as stacked histograms for the DDFs: COSMOS (blue), ELAIS (purple), DECaPS-East (orange), and DECaPS-West (green).
\label{fig:3}}
\end{figure}

The strategy for the cadence of this DDF program is very important for our science goals.
As described in \S~\ref{ssec:survey_fieldsel}, series of images were obtained in multiple 20-30 minute sequences during the night, and this was repeated every $\sim$3 nights.
Figure~\ref{fig:3} shows the time between images in a series (top); the time between sequences within a night (intra-night gaps, middle); and the time between observing nights (inter-night gaps, bottom).
The top panel of Figure~\ref{fig:3}, the inter-image time, is essentially the distribution of image readout times (i.e., unavoidable overhead). 
The distribution is centered on $\sim$29 seconds, as expected for DECam.

For our program, when possible, we attempted to spread the 20-30 minute sequences out during the night in order to have a better chance of detecting objects that rise or fade within hours.
The middle panel of Figure~\ref{fig:3} shows how we were only able to schedule inter-sequence time gaps regularly for the COSMOS field in 2021A, when $\sim$60\% of sequences separated by $>$10 minutes.
It also shows that on $\sim$8 nights we were able to schedule an inter-sequence gap for the DECaPS-East field ($\sim$15\% of nights).
During 2021B, the two primary DDFs (ELAIS and DECaPS-West) always had their sequences done back-to-back for scheduling convenience, and so do not appear at all in the middle panel of Figure~\ref{fig:3}.
Furthermore, when a field is rising or setting, we avoid inter-sequence time gaps in order to minimize the airmass of the observations and improve image quality.

The bottom panel of Figure~\ref{fig:3} demonstrates that the targeted inter-night cadence of $\sim$3 days is achieved for most fields, most of the time.

\subsubsection{Bright-time Observations}\label{sssec:survey_char_bright}

\begin{figure*}
\includegraphics[width=\columnwidth]{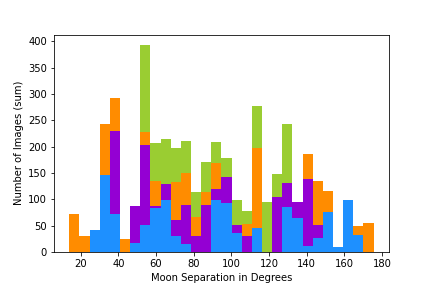}
\includegraphics[width=\columnwidth]{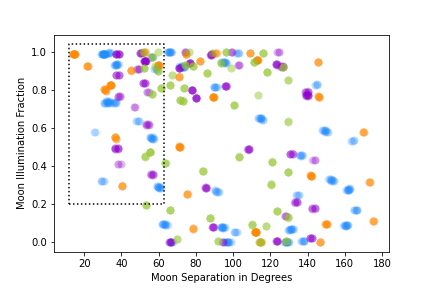}
\includegraphics[width=\columnwidth]{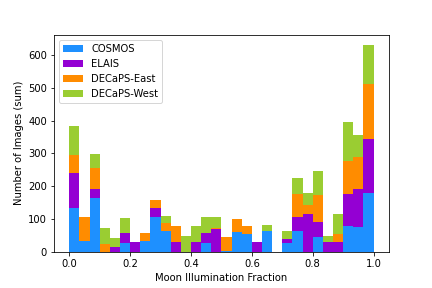}
\includegraphics[width=\columnwidth]{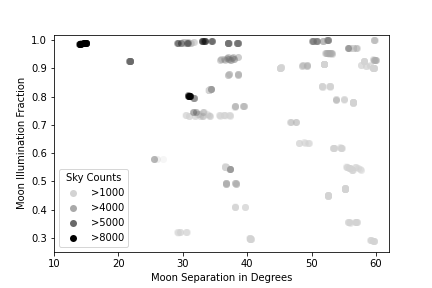}
\caption{
\textit{Left:} Stacked histograms of moon angle in degrees (\textit{top}) and moon illumination fraction (\textit{bottom}) at the time of our observations.
Colors represent the four fields, COSMOS (blue), ELAIS (purple), DECaPS-East (orange), and DECaPS-West (green), as shown in the legend in the lower-left panel.
\textit{Right:} Scatter plots of moon separation \textit{vs.} illumination.
In the top panel, points are colored by field, and a dotted-line box represents the "region of concern" in which we investigate sky background levels (\textit{lower-right}).
In the bottom panel, the point shading is representative of the sky background in counts (as in the legend), for DDF images obtained within $\leq$60 degrees of the moon, at a time when the moon illumination factor was $\geq$0.2.
A sky background of $5000$ counts is the targeted limit for our DECam images (for any filter).
\label{fig:4}}
\end{figure*}

Our original proposed DECam DDF program avoided observing during bright time, but the DECAT programs were co-scheduled by CTIO for a three night cadence without moon avoidance.
Figure~\ref{fig:4} shows the distribution of moon separation and moon illumination for all of the images (left panels), as well as the distribution of the images in separation \textit{vs.} illumination (upper-right panel). 
Since a sizeable fraction of the images were obtained with low moon separation and high moon illumination, we can correlate these moon parameters with the sky background rate and investigate observing strategies to keep the total background $\lesssim$5000 counts.
In the upper-right panel of Figure~\ref{fig:4} we defined a "region of concern" of moon separation $\leq$60 degrees and moon illumination fraction $\geq$0.2 (dotted box).

For all images obtained within the "region of concern" we plot their sky background in counts as a function of moon separation and illumination in the bottom-right panel of Figure~\ref{fig:4}.
This plot shows that in only a few cases does the sky background exceed 5,000 counts for image (in any filter) obtained with a sky background of $\lesssim$40 degrees and a moon illumination fraction of $\gtrsim$0.8. 
This correlation was something we realized by the end of the 2021A semester, and in 2021B implemented a more aggressive moon avoidance strategy.
Skipping nights with a bright, nearby moon caused the 2021B fields, ELAIS and DECaPS-West, to more often have an inter-night time gap of 6 days (purple and green histograms in the bottom panel of Figure~\ref{fig:3}) but fewer instances of low moon separation (purple and green points in the upper-right panel of Figure~\ref{fig:4}).
In Section \ref{ssec:proc_imchar} we further explore the effect of high sky background on our ability to detect transients and variables.

\section{Data Processing and Characterization}\label{sec:proc}

When this DECam DDF program began in 2021-A, in order to get started immediately we made the decision to adopt a well-established pipeline with demonstrated success: the real-time automated difference imaging pipeline which was originally developed to rapidly discover new optical transients in gravitational wave follow-up imaging with DECam \citep{2019ApJ...881L...7G}.
As one of the main technical goals of this program is to "enrich our current time-domain alert ecosystem", the difference-image detections are used to generate and distribute alert packets using the same conventions as the ZTF, which is an early version of the LSST architecture \citep[e.g.,][]{2019PASP..131a8001P}.
Our implementation of these data processing and alert generation pipelines is described below in Sections~\ref{ssec:proc_egal} to \ref{ssec:proc_alerts}, and the processed images and difference-image detections (which trigger alerts) are characterized in Sections~\ref{ssec:proc_imchar} and \ref{ssec:proc_srcdet}.
The long-term goal for the DECam DDF processing pipeline is to use the LSST Science Pipelines\footnote{\url{https://pipelines.lsst.io}}, which are currently under active development.

All of the processing described in this work is run at the National Energy Research Scientific Computing Center (NERSC).
Most of the image and catalog data products were not publicly available at the time of this work's preparation, with a few exceptions.
The raw and reduced\footnote{All raw DECam data is processed by a Community Pipeline managed by NOIRLab, as described in the DECam Data Handbook, \url{https://noirlab.edu/science/documents/scidoc0436}.} images from this DDF program are available without a proprietary period via the NOIRLab archive.
Alert packets are distributed to a variety of brokers, as described in Section\ref{ssec:proc_alerts}, and a vetted subset of 4413 "probably-real" candidates and their light curves are made available via GitHub for easy access, as described in Section~\ref{sssec:proc_canddet_github}.
The long-term goal for the DECam DDF data products is to also publicly release, e.g., processed images, templates, nightly stacks, deep stacks, and catalogs which include forced photometry, as listed in Section \ref{sec:conc}.

At the time of this work's preparation, efforts towards these goals of using the LSST Science Pipelines and making more data products available from the DECam DDF were underway.

\subsection{Image Reduction and Difference-Image Source Detection Pipeline}\label{ssec:proc_egal}

The pipeline searches for transient and variable objects by performing image subtraction of the science images, using either manually constructed template images or template images built from pre-existing surveys.
It then identifies residuals on the difference images.
The pipeline is modified from a pipeline for finding transients in DECam images originally written by D. Goldstein \citep{2019ApJ...881L...7G}.

The pipeline begins by ingesting raw images directly from the NOIRLab data archive, dividing the image stack into individual images for each chip (a total of 60 images; we remove two bad chips, CCDS 31 and 61\footnote{\url{https://noirlab.edu/science/programs/ctio/instruments/Dark-Energy-Camera/Status-DECam-CCDs}}).
It performs preliminary standard data reduction steps: overscan and bias correction, flatfielding using standard observatory flatfield frames, and a linearity correction using the observatory-supplied lookup table of device counts and linearity corrected counts.
It flags all pixels brighter than 90\% of the saturation level in the header, and it also flags pixels mirrored across the long center line of the image to avoid any contamination from amplifier crosstalk.

{\tt Source Extractor} \citep{1996A&AS..117..393B} is used to detect and measure all sources on the image, and these sources are then used to estimate the seeing of the image and in astrometric and photometric calibration.
The pipeline calibrates each chip's image astronomy, stored in the image header as an updated world coordinate system (WCS), by running {\tt SCAMP} \citep{2006ASPC..351..112B} to match objects identified on the image with stars drawn from the Gaia DR2 catalog \citep{2018A&A...616A...1G} using the NOIRLab datalab Query Client\footnote{\url{www.github.com/astro-datalab/datalab}} \citep{2014SPIE.9149E..1TF}.  The pipeline then uses {\tt SWarp} \citep{2002ASPC..281..228B} to solve for the WCS using these matched objects.

To determine image zeropoints, the pipeline searches
the Dark Energy Survey (DES) \citep{2021ApJS..255...20A},
the Dark Energy Camera Legacy Survey (DECaLS) \citep{2019AJ....157..168D},
the DECam Plane Survey (DECaPS) \citep{2018ApJS..234...39S},
and the PanSTARRS \citep{2016arXiv161205560C} catalogs for stars in the field to use as photometric calibrators.  Searches of DES, DECaLS, and DECaLS use the NOIRlab Query Client; searches of PanSTARRS use the Vizier catalog services \citep{2000A&AS..143...23O}.  
In the case of PanSTARRS, it transforms the catalog magnitudes to the system used in the DECam $g$, $r$, and $i$ filters\footnote{Ryan Ridden-Harper, private communication}; the other surveys were performed using DECam, so no photometric transformations are necessary.
Objects found in the science images are matched to this list of photometric calibrators using SCAMP, and the median of the measured FWHM of the matched objects on the science image is saved as the image's seeing.
The number of counts on these objects is measured in an aperture whose radius is 0.6731 times this seeing (chosen to match the aperture that will be used for science measurements later), and a zeropoint for the image is determined from these measurements and the corresponding catalog magnitudes.

Next, the pipeline identifies template images (see Section~\ref{ssec:proc_templates}).
It generates an object catalog for the base template images with {\tt Source Extractor}, and then uses {\tt SCAMP} and {\tt SWARP} to align the template image with the science image.
It subtracts the template image from the science image using the {\tt HOTPANTS} package \citep{2015ascl.soft04004B}\footnote{For future analyses, we plan to also implement both the ZOGY (\citep{Zackay_2016} and the Saccadic Fast Fourier Transform ({\tt SFFT}) \citep{2021arXiv210909334H} algorithms for image differencing .}.
It creates a noise image for the subtraction image by adding the science noise image to a rescaled warped-reference noise image in quadrature; the reference noise image is scaled by the relative normalization of the two images found by {\tt HOTPANTS}.
This is an approximation to the true noise in the subtraction.
Because the reference image was resampled and convolved, there exist correlations between the pixels in the template image; our reported photometry does not take these correlations into account.
However, as noted in Section~\ref{ssec:proc_templates}, the template images are all enough deeper than the science images that the noise in the difference image is dominated by the noise in the science image, meaning that any correlations between pixels are not significant.
Because the template images are so much deeper than the science images, we configure the subtraction to always perform seeing-matching convolution on the reference image, even in occasional case where the science image has better seeing; {\tt HOTPANTS} is able to handle this situation.
While we haven't performed detailed comparisons, from a few trials we observed that artifacts from "backwards" convolution are not as severe as the complications that arise from the correlated pixel noise that results from convolving the noisier science image.

The pipeline finally runs {\tt Source Extractor} on the resultant difference image to identify residual signals.
Most of the signals detected are in fact artifacts.
How the pipeline tries to identify these artifacts, and which signals are used to generate alerts, is covered in Sections~\ref{ssec:proc_RB} and \ref{ssec:proc_alerts}.
Once sources have been identified, the pipeline measures fluxes on difference images again with {\tt Source Extractor}, this time using it to do forced photometry at the position of the detected source.
It uses the difference noise image described above to estimate the uncertainties on these fluxes.
The reported fluxes are in circular apertures with a radius equal to 0.6731 times the FWHM of the seeing on the science image.
These fluxes are in arbitrary units, but the image zeropoint included in each alert was determined using the same aperture, so it effectively provides both the aperture corrections and the units of the flux measurements.
The pipeline does not currently build and search end-of-night stacks, but that is functionality that will be implemented in the future.

\subsection{Template Images}\label{ssec:proc_templates}

The pipeline is designed to automatically build subtraction templates when it processes a new image.
It maintains a cache of template images (one image for a single chip from the detector) for fields it has seen previously.
If the image it is processing has at least 90\% overlap with a cached template, it will use that template image.
(At the moment, there is only one cached template for any given field.
In the future, we hope to add the ability to add newer, higher-quality templates, and the pipeline will then select the deepest and/or best seeing template for use.)
If there is no existing template, the pipeline will automatically search the images from the DES DR1, DECaLS DR9, and DECaPS DR1 surveys.
(In practice, we did not use templates from DECaPS; see below.)
It will download the coadded images from these survey's data releases, and will stitch them together to make a template for each chip of the exposure being processed.
The combination is performed using SCAMP to align the images and SWARP to add them together.
This process results mostly in juxtaposing the survey images, but will coadd and scale the small regions of overlap.
The resulting template will then be saved to the template cache for future use.
In the case where no template is available in the surveys the pipeline knows how to search, one may manually build a template and add that to the template cache.

In practice, for the extragalactic $g$ and $r$-band fields (both COSMOS and ELAIS), the pipeline built templates from the DECaLS survey images (DECaLS did not observe in $i$-band).
For the ELAIS $i$-band fields, the pipeline built templates from the DES survey images.
As COSMOS was not in the DES footprint, we manually built an $i$-band template using pre-existing publicly available images of the same field that we obtained from the NOIRLab archive, chosen to have the best available seeing and depth.
For the galactic fields, even though survey templates were available, we built manual templates.
Because the archived survey images are not aligned exactly with our fields, two or more survey images must be stitched together to provide a template overlapping the science field.
This leads to a spatial discontinuity in the PSF at locations on the template that are on the border between different source archive images.
In practice, this was not a serious problem for extragalactic fields.
However, galactic fields are so crowded with stars that this led to a very large number of artifacts as a result of the failure of the difference imaging software to handle this spatial PSF variation.
As such, for the galactic fields, we built templates by coadding all of the images from one night early in the survey that had stable zeropoints and good seeing; for DECaPS-East, we used the images from the night of 2021-04-14, and for DECaPS-West, we used the images from the night of 2022-01-12.
For all of the templates we built manually, we used SCAMP to align the images, and SWARP to build the image stack.
We used the "clipped-mean" algorithm in SWARP in order to reject cosmic rays and other artifacts present in individual images.

The template images used for the present survey are significantly deeper than the images that were taken as part of the survey.
For the images built from the DECaLS image stacks, the effective level of the sky noise is two orders of magnitude lower than sky noise in our best science images.
For the i-band extragalactic fields, the effective sky noise was a factor of 5--10 lower in the template than in the science images.
For the galactic fields, the template sky noise was a factor of a few lower.
(This will become a limiting factor in the future when we analyze full-night stacks, requiring a new, deeper template for the galactic fields.)
In all cases, the noise in the difference images are dominated by the noise in the science images.
One consequence of using images from a survey data release for a template, as opposed to manually building templates specific images chosen for their quality, is worse seeing.
The templates for images from the surveys have a typical seeing of $\sim1.4"$, as compared to the $\sim1.0"$ seeings for the manually-built templates.

\subsection{Analytic Residual Cuts and the Real/Bogus Classifier}\label{ssec:proc_RB}

The end goal of the pipeline is to find transient and variable sources in the images, including both true transients such as supernovae and objects that brightened relative to the template image; it does this by detecting positive residuals on the difference images.
(Currently, the pipeline only detects positive signals on the difference image, and as such will not find variable objects where the object was dimmer in the search image than it was in the template image.)
The difference images have numerous artifacts in the data, requiring further work to improve the quality of the set of detected sources.

The pipeline makes a few basic analytic cuts based on the detected residuals for the parameters measured by {\tt Source Extractor}.
Any objects that include a flagged (saturated or bad) pixel are rejected.
Further cuts reject objects with a FWHM more than twice the seeing, objects whose major and minor axes have a ratio larger than 1.5, objects whose major axis is less than 1 pixel, objects whose average pixel flux uncertainty in a 6-pixel radius aperture is more than 1.25 times the median image pixel uncertainty, objects with S/N$<5$, and objects within 10 pixels of the edge of the image.
A final cut tries to eliminate "dipoles" by rejecting objects that have too many negative pixels (resulting from a small misalignment or convolved PSF mismatch).
This cut looks at pixels in a square box that is 4$\times$FWHM on a side, identifying all pixels that are below zero by $>2\sigma$ and all pixels that are above zero by $>2\sigma$.
If either the number of negative pixels is more than half the number of positive pixels, or the absolute value of the sum of the flux in the negative pixels is more than half of the sum of the flux in the positive pixels, the object is rejected.
Although these analytic cuts reduce the number of artifacts, after they have been applied, many artifacts remain.

To further filter the catalog of difference-image source detections without the prohibitive effort of manual scanning all of them, the pipeline uses an automated scanning of candidates using a machine learning (ML) system similar to that described in \citet{2015AJ....150...82G} to produce a "real/bogus" (R/B) score for each candidate.
It starts with $51\times 51$ pixel cut-outs of the science, template, and difference images centered on each remaining residual, scaled to greyscale images using
the {\tt ZScaleInterval} module in the {\tt Astropy Visualization} package.
These image triplets are passed to a convolutional neural network \citep{Ayyar_CNN}, which returns an R/B score for each candidate.
Ideally, R/B values near 1 indicate that the detection is probably an astronomical point source (i.e., real), and values near 0 indicate the detection is probably an artifact of the camera or reduction pipeline (i.e., bogus).

For the images acquired and alerts released in both the Spring and Fall semesters of 2021, we used the ML model trained in \citet{Ayyar_CNN} using images from \citet{2015AJ....150...82G}.
However, this model was not trained on a dataset identical to our program: although the images were from DECam, the positive detections were based on simulated point sources injected into images.
To judge how well the simulation-trained ML model was able to reproduce manual vetting of actual detected sources, we did some additional manual vetting of detections from our processing as described in \citet{Ayyar_CNN}.
By using these vetted candidates as a new training set for the ML model, we found that we could improve its performance.
In particular, while the rate of false positives was similar at a few percent, we were able to reduce the missed detection rate from $\sim1/2$ to about 5\% for extragalactic fields.  The observers who performed the manual vetting of candidate detections were given a randomly chosen sample of all of the detections the pipeline had produced, with the goal of producing a representative sample of detections.
One side effect of this is that rarer, brighter candidates were not well-represented in the training sample, likely limiting the quality of the vetting for the brighter candidates; a possible implication of this is discussed in Section~\ref{ssec:proc_srcdet}.
We plan further work on retraining the ML system to better represent candidate detections of all magnitudes.

Thus, we have used these new R/B scores for the analysis in this paper.
We show the distribution of R/B values and describe the cutoffs that we apply to identify likely-real phenomena for analysis in Section~\ref{ssec:proc_srcdet}.
Manual vetting for the Galactic fields was still ongoing at the time of publication, which is partly why this paper focuses on detections in the extragalactic fields.
In future semesters, we will use this retrained model for alert generation and distribution.
Anyone using the DECam DDF alerts would find the new R/B scores in the alert packets, and any user in need of more detailed information about the retraining should please reach out to the authors of this work.

\subsection{Alert Packet Creation and Schema}\label{ssec:proc_alerts}

The pipeline sends an alert packet (in Apache Avro\footnote{\url{https://avro.apache.org/}} format) for every detected residual that passes the R/B cutoff (which varies based on the ML model in use).
The schema for each alert includes the Right Ascension (RA) and Declination (Dec) of the object, as well as some type and photometric redshift information from the Legacy Survey DR9\footnote{\url{https://www.legacysurvey.org/dr9/}} \citep{2019AJ....157..168D} if there is a nearby object in that survey.
The "host" determination is a very simplistic search; it simply returns the closest object within 10\arcsec of the detected residual.
Note that because the templates used in this pipeline may well not be the images used to build the Legacy Survey catalog, and because the DR9 catalogs include fluxes that have been through a deblending procedure, any flux information from that catalog is \emph{not} directly useful for combining with difference flux measurements in the alert (for instance, to try to reconstruct the total flux of the detected residual plus any underlying host light).
These fields will be null in the alert for any source outside of the DR9 footprint (which includes all of the Galactic fields), or that has no DR9 object within 10\arcsec.
The alerts also include parallax information from Gaia if the object overlaps a Gaia star.
(For practical reasons, the pipeline does not actually search for this information for Galactic sources.)

Matching the LSST and ZTF conventions, each alert includes records corresponding to the triggering detection as well as a history of prior detections in difference images.
These records include information about the residual detection's RA, Dec, flux, magnitude, FWHM, and R/B score\footnote{Full schema for the alerts may be found at \url{www.github.com/rknop/decat_schema}.}.
It also includes some information about the image on which the candidate was detected (MJD of the exposure, integration time, filter band, seeing, zeropoint, sky variance).
Finally, the alert includes 51$\times$51-pixel ($\sim$13$\times$13$\arcsec$) JPEG cutouts of the science, template, and difference images from the detection which triggered the alert.

The goal is to provide alerts in as close to real-time as possible.
Currently, the pipeline is able to produce alerts for extragalactic fields within 10 minutes of the raw data becoming available at NOIRLab (which happens shortly after the observation is complete).
For galactic fields, which take longer to process (as both sky subtraction and object extraction take longer in crowded fields), the time to generate alerts is 30--60 minutes.
For most of this work, which was in development mode, we were not actively running the pipeline as images were coming in, but would run a night's worth of data the next morning when we could monitor the process.

Alerts are sent to the Kafka-based alert distribution system hosted by the University of Washington and also used for ZTF (the ZTF Alert Distribution System, or ZADS; \citealp{2019PASP..131a8001P}).
Community alert brokers may then stream DECam DDF alerts from the ZADS system as they become available.
Brokers connected to the ZADS system include Alerce \citep{2021AJ....161..242F}, ANTARES \citep{2021AJ....161..107M}, AMPEL \citep{2019A&A...631A.147N}, Fink \citep{2021MNRAS.501.3272M}, Lasair \citep{2019RNAAS...3...26S}, and Pitt/Google.  
Alerts use a topic structure of  \texttt{decat\_caldate\_propid}, where \emph{caldate} is the calendar date of the night in which the science image was taken, and \emph{propid} is the NOIRLab proposal ID under which the science image was taken.

\subsection{Processed Image Characterization}\label{ssec:proc_imchar}

\begin{figure*}
\includegraphics[width=17cm]{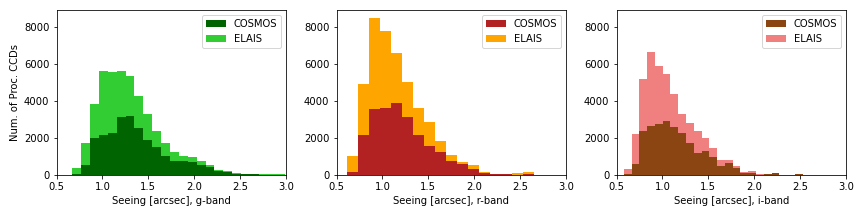}
\includegraphics[width=17cm]{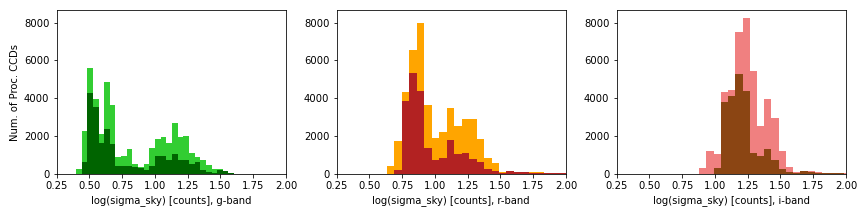}
\includegraphics[width=17cm]{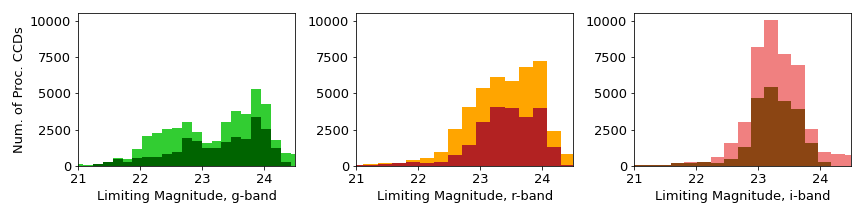}
\caption{Stacked histograms of the seeing (\textit{top}), $\sigma_{\rm sky}$ (sigma\_sky; \textit{middle}), and limiting magnitude (\textit{bottom}), in filters $g$, $r$, and $i$ (\textit{left to right}), as measured in the processed CCDs for the extragalactic fields, COSMOS and ELAIS. \label{fig:5}}
\end{figure*}

In Section \ref{ssec:survey_char} we characterized the survey, the conditions in which the images were obtained, and explored the impact of moon illumination and proximity on the sky background in the raw images.
In order to characterize the processed images for the extragalactic deep fields (COSMOS and ELAIS), in Figure~\ref{fig:5} we show the distributions of the limiting magnitude ($5\sigma$ detection limit); seeing (full-width half-max of the point-spread function, in arcseconds); and sky background, $\sigma_{\rm sky}$ (the median absolute residual of pixel flux, $f$, from the median pixel flux, $|f-\tilde{f}|$, in counts), as measured in the individual processed CCD images.

The top row of Figure~\ref{fig:5} shows the seeing distribution, and how the majority of observations were obtained with $<$1.5\arcsec seeing, with a tail to poorer image quality.
Although there is a correlation between seeing and airmass (not shown in a plot), we find that the higher-airmass exposures (Figure~\ref{fig:2}) is not solely responsible for the poor-seeing tail.
The middle row, which contains the histograms for $\sigma_{\rm sky}$ (labeled as ${\rm log(sigma\_sky)}$), shows that a minor fraction of the COSMOS observations have a large sky background, and that a significant fraction of the ELAIS observations experience a high sky background.
The bottom row of Figure~\ref{fig:5} shows that this program is often reaching the anticipated single-image limiting magnitude depths ($r\sim23.5$ mag) for the COSMOS and ELAIS fields -- but not always.

\begin{figure*}
\includegraphics[width=17cm]{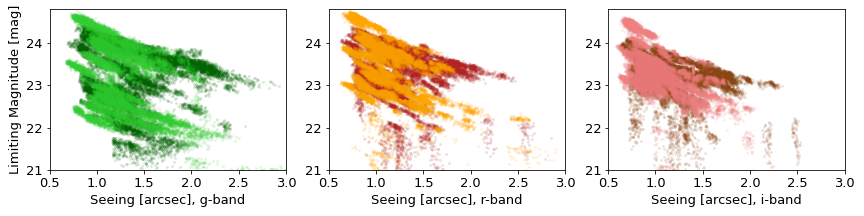}
\includegraphics[width=17cm]{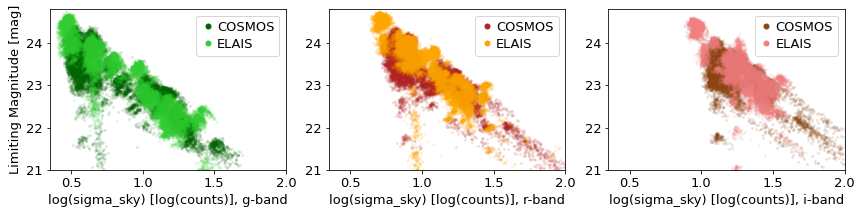}
\includegraphics[width=17cm]{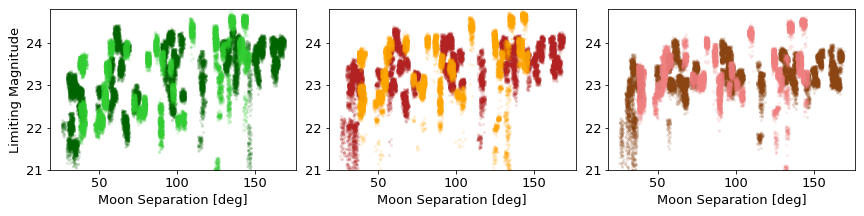}
\includegraphics[width=17cm]{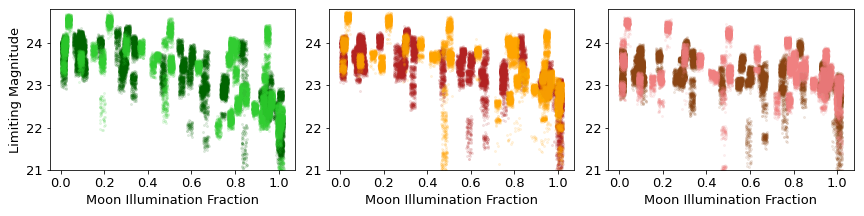}
\caption{The image seeing (\textit{top row}), sky background $\sigma_{\rm sky}$ (sigma\_sky, \textit{second row}), moon separation (\textit{third row}), and moon illumination (\textit{bottom row}) versus the limiting magnitude for all processed DECam CCDs, for filters $g$, $r$, and $i$ (\textit{left to right}), for the extragalactic fields COSMOS and ELAIS.
This shows how image depth is correlated with image quality (IQ; the seeing) and sky background, due to the underlying impact of moon separation and illumination.
These correlations are stronger for $g$-band than for $i$-band.
A small random scatter has been added to the moon separation and illumination values to better see the points.
Each point represents one CCD. \label{fig:6}}
\end{figure*}

In Figure~\ref{fig:6} we show how the limiting magnitude is correlated with seeing (top row), sky background (second row), moon separation (third row), and moon illumination fraction (bottom row) for all of the processed CCDs for COSMOS and ELAIS.
As expected, the limiting magnitude is correlated primarily with sky background (second row), and secondarily with seeing (top row). 
The correlations with moon separation and illumination are also clear; illumination appears to have a stronger impact, as expected.
Figure~\ref{fig:6} shows that, compared to the COSMOS field, the ELAIS field received relatively fewer observations with large moon separations and more with high moon illumination fraction.
This lead to the ELAIS field having relatively more images with high sky background and, correspondingly, brighter limiting magnitudes.
In Section \ref{ssec:proc_srcdet} we explore how these factors affect the images' limiting magnitudes, and the impact on our source detection capabilities in the extragalactic fields.

\subsection{Difference-Image Object Characterization}\label{ssec:proc_srcdet}

In this section and the next, Section~\ref{ssec:proc_canddet}, we use the catalogs of detected and associated difference-image sources, from which the alert packets are generated, for our analysis.
We use these catalogs because the alert packets are sent to brokers but not persisted at NERSC, where the pipeline runs and where the catalogs are stored permanently, and because the JupyterLab hosted by NERSC is convenient for collaborative analysis.
Thus, for this work, we use terminology matching the catalogs, not the alert packet schema, which we have adapted to be similar to ZTF/LSST alerts.
We will refer to {\it objects} as a single detection in a difference image, and {\it candidates} as the set of associated objects at a given sky coordinate.
This section focuses on characterizing objects, and then the next section focuses on characterizing candidates.

\begin{figure*}
\includegraphics[width=17cm]{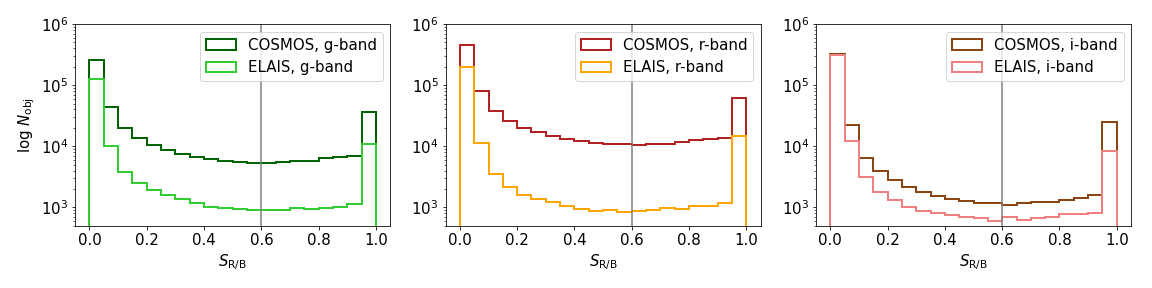}
\includegraphics[width=17cm]{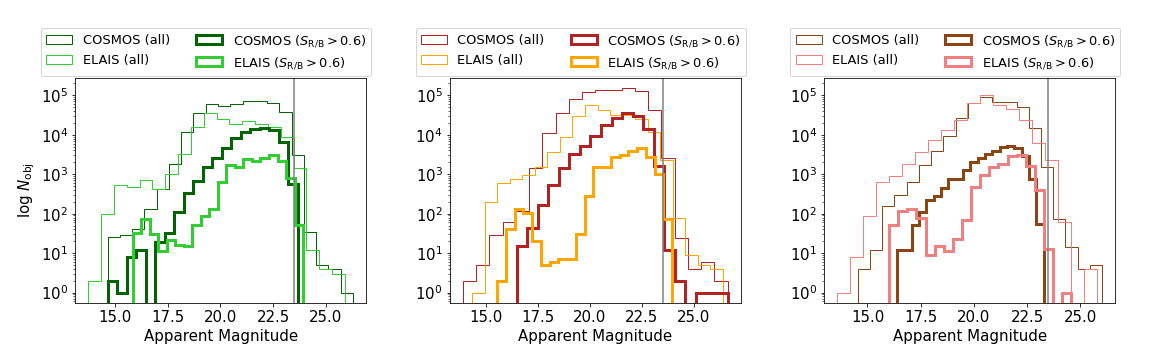}
\caption{Characterizing the objects detected in the DECam difference images (where an "object" is a source in a difference image, and $N_{\rm obj}$ is the number of objects per difference image).
{\it Top:} Histograms of object R/B scores ($S_{\rm R/B}$) for the COSMOS and ELAIS fields, for $g$, $r$, and $i$-band images (left to right).
A grey line is shown at $S_{\rm R/B}=0.6$, the minimum of the histogram in $g$-band.
{\it Bottom:} Histograms of object apparent magnitudes for each field and filter (as in the top panel), with the thicker lines representing the distribution when only objects with $S_{\rm R/B}>0.6$ are included.
A grey line is shown at $m=23.5$ mag for reference.
\label{fig:7}}
\end{figure*}

The two main properties of interest for objects are their apparent magnitude and their real/bogus score ($S_{\rm R/B}$), the latter of which is assigned using a machine learning algorithm as described in Section \ref{ssec:proc_RB}.
Figure~\ref{fig:7} shows the distributions of these properties for all objects in the extragalactic fields.
Recall that COSMOS will have more objects than ELAIS despite having a similar number of epochs in 2021 because the COSMOS field had three DECam pointings, but the ELAIS field had only two DECam pointings.
COSMOS is also at a lower Galactic latitude ($b=42$ deg) than ELAIS ($b=-73$ deg), and so has more stars, and thus more variable stars and more difference-image objects.

The top row of Figure~\ref{fig:7} demonstrates how there is a minimum in the distribution of R/B scores at 0.6 (grey line), suggesting that most real astrophysical sources have $S_{\rm R/B}>0.6$ and most artifacts have $S_{\rm R/B}<0.6$.
The bottom row shows how the shape of the apparent magnitude distribution changes when only objects with $S_{\rm R/B}>0.6$ are included.
For example, in the $g$-band we can see that most of the brightest and faintest objects are likely artifacts with $S_{\rm R/B}<0.6$.
We can also see that the number of objects detected turns over at about $\sim$22.5 mag, and drops steeply beyond $\sim$23.5 mag (grey line), which is a little brighter than the direct image limiting magnitudes discussed in Section \ref{ssec:proc_imchar}.
This is to be expected, as some signal is inevitably lost in the difference-image processing.

In the bottom row of Figure~\ref{fig:7}, for the ELAIS field we see a bimodal distribution with a bright-end component that peaks at $\sim17$ mag in all filters, which looks suspiciously pronounced with a logarithmic $y$-axis.
Multiple factors contribute to this.
Further investigation (not shown in a plot) reveals these couple hundred objects with $S_{\rm R/B}>0.6$ to be very bright stars near the saturation limit, part of a population that can be identified and isolated by their inverse correlation between difference-image brightness and R/B score, and which could be removed from the population during scientific analysis.
For now, we leave them in the set of objects in order to fully characterize all detections for the reader.
This population is not altogether absent from the COSMOS field, but is not as conspicuous for a confluence of factors--- COSMOS has more stars overall; ELAIS on average has better seeing (see Figure~\ref{fig:5}, top row); differences in template depths and image quality; etc.
Second, as discussed in Section~\ref{ssec:proc_RB}, the R/B classifier was trained on a representative sample of detections, meaning that the number of brighter sources (magnitude$\lesssim 20$) in the training sample was small.
Empirically, the R/B classifier passes a smaller fraction of candidates in the magnitude range 18--20 than it does outside that range.
This suppresses the number of detections in that magnitude range for both COSMOS and ELAIS; the greater number of object detections at brighter magnitudes in ELAIS combined with this effect leads to the observed dip in the bottom row of Figure~\ref{fig:7}.

\begin{figure*}
\includegraphics[width=17cm]{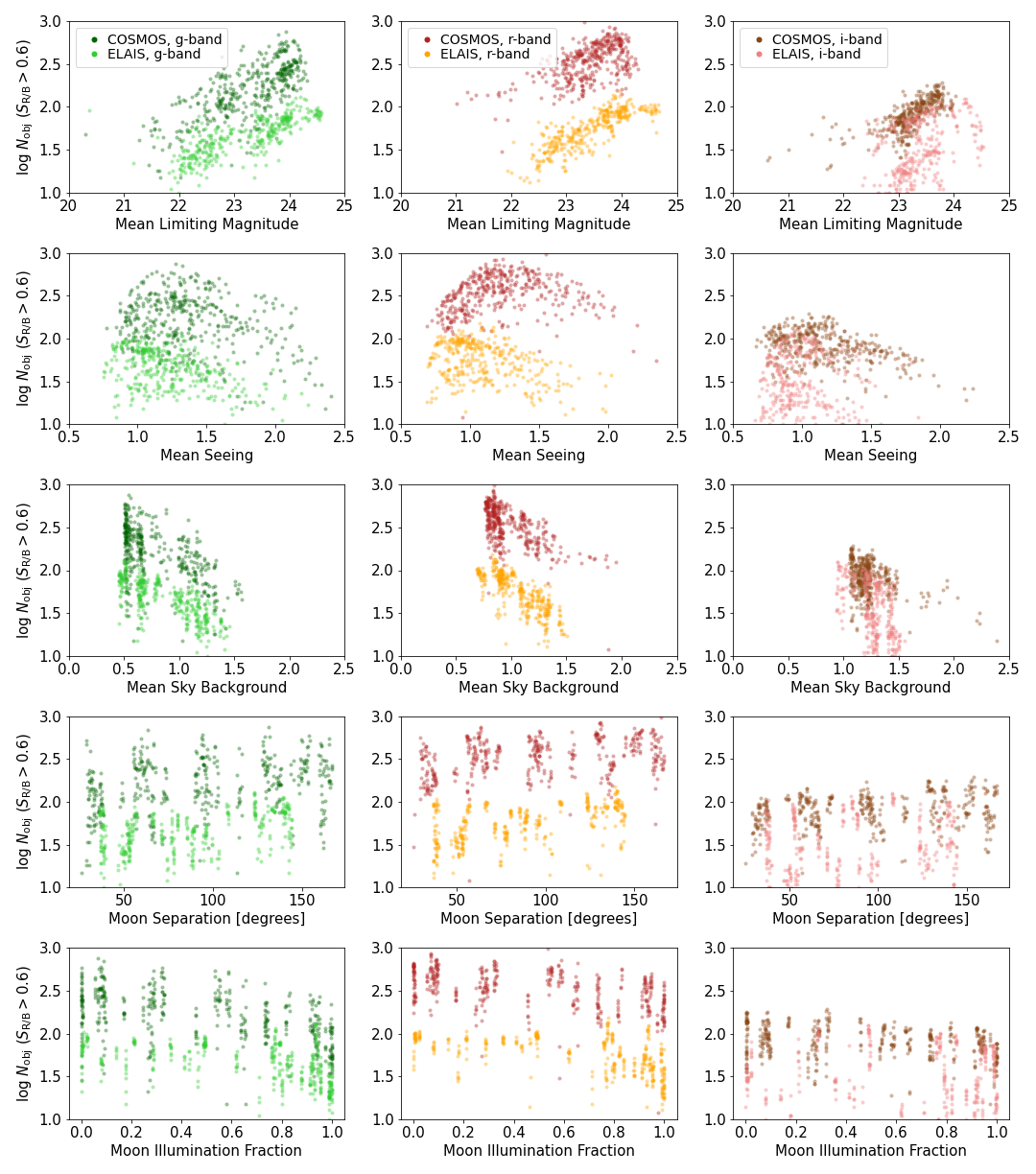}
\caption{The number of objects ($N_{\rm obj}$) with R/B score $S_{\rm R/B}>0.6$ that were detected in the difference image for exposures of the extragalactic DDF COSMOS and ELAIS in $g$, $r$, and $i$-bands (left to right) versus the mean image quality parameters (an average over all CCDs) of limiting magnitude (top row), seeing (second row), and sky background (third row).
The fourth and fifth rows show how the number of objects varies with moon separation and illumination fraction.
The COSMOS fields have more difference-image objects than the ELAIS fields because they cover more area and are at lower Galactic latitude.
\label{fig:8}}
\end{figure*}

In Section \ref{ssec:proc_imchar} we characterized the images obtained by this program and showed how the limiting magnitude was affected by the seeing and sky background.
Figure~\ref{fig:8} shows how the number of objects per difference image ($N_{\rm obj}$) detected with $S_{\rm R/B}>0.6$ varies with image quality parameters limiting magnitude, seeing, and sky background, and with the moon separation and illumination at the time of the observation.
As described in Section \ref{ssec:proc_egal}, the pipeline processes images and calculates the image quality parameters for each individual CCD; in Figure~\ref{fig:8} we use the total number of objects detected in all CCDs, and the mean values of the limiting magnitude, seeing, and sky background over all CCDs.

Figure~\ref{fig:8} demonstrates how the number of objects with $S_{\rm R/B}>0.6$ detected in a difference image is correlated with all of the image quality parameters and moon conditions in the $g$ and $r$-bands, but less so in $i$-band.
We can also see that the number of objects {\it per exposure} is higher for the COSMOS field than the ELAIS field; as previously mentioned, this is likely the result of a higher stellar density in COSMOS.
In the second row of Figure~\ref{fig:8} we can see that the number of objects detected with $S_{\rm R/B}>0.6$ does not decrease steadily as the seeing increases, as expected, and that a peak at a seeing of $1.1-1.4\arcsec$ is particularly pronounced for the COSMOS images.
This is, in part, an indication of what we already know: that imposing a cut of $S_{\rm R/B}>0.6$ increases the purity of the object sample, but does not perfect it.
Further investigation (not shown in a plot) has revealed that the location of this peak coincides with the PSF FWHM for the template images ($1.1-1.4\arcsec$ for the three COSMOS fields, and $\sim1.0\arcsec$ for the two ELAIS fields), and furthermore that there are more objects with poorer R/B scores ($0.6 < S_{\rm R/B} < 0.9$) detected in difference images with seeing of $1.2-1.4\arcsec$, which contribute to the peak in the number of objects versus seeing.
Together, these facts reinforce that future studies are needed to characterize the R/B scores as a function of image quality, and this remains a goal for us (Section~\ref{sec:conc}).
For this work, we continue to use a cut of $S_{\rm R/B}>0.6$ to increase the purity of the sample set as we characterize detections.
The bottom two rows of Figure~\ref{fig:8} show that the sky background due to moonlight does severely impact the detection rate in the $g$-band, but does not reduce it to zero.
This tells us that continuing to observe through bright time is not a useless endeavor.

Although not shown in this work, a version of Figure~\ref{fig:8} made for the number of objects with $S_{\rm R/B} < 0.6$ per exposure (i.e., objects \textit{less} likely to be real) show much larger $N_{\rm obj}$ values and weaker correlations with image quality.
The latter indicates that poorer image quality is not the primary cause of low-R/B objects, but rather the causes are persistent or systematic (e.g., bright stars; detector artifacts; processing pipeline flaws; R/B training set), and thus that with future work they can be characterized and removed from the sample.

\subsection{Candidate Characterization}\label{ssec:proc_canddet}

\begin{figure*}
\includegraphics[width=17cm]{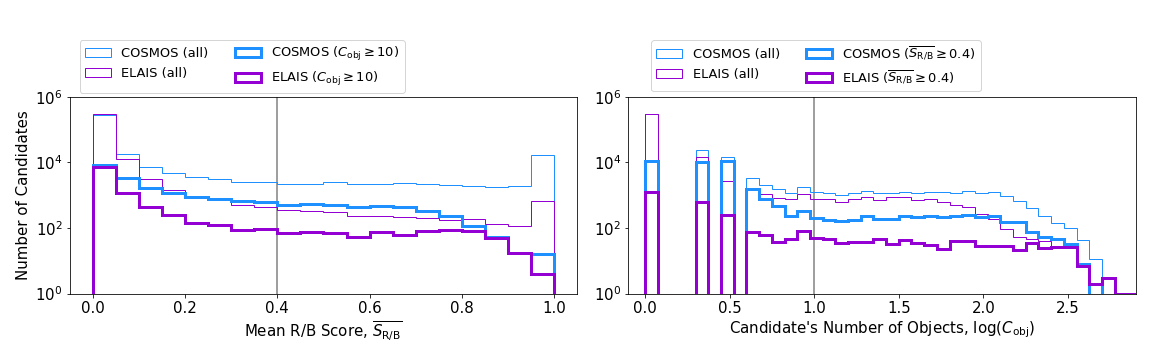}
\includegraphics[width=17cm]{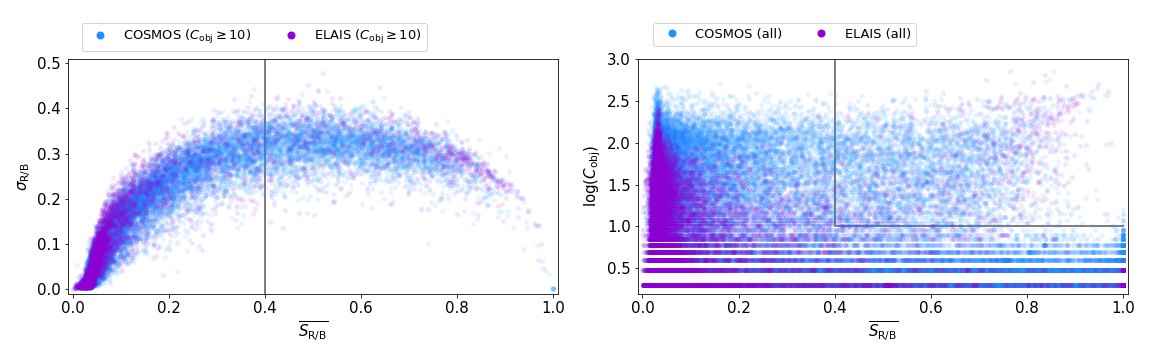}
\caption{Characterizing the candidates (where candidates are objects associated by sky coordinate) in the COSMOS (blue) and ELAIS (purple) fields.
Grey lines mark the limits of $C_{\rm obj}>10$ and $\overline{S_{\rm R/B}}>0.4$ that are applied to identify potentially real candidates of interest.
{\it Top left:} The distribution of mean R/B score for all candidates (thin lines), and for candidates with $>$10 objects (thick lines).
{\it Top right:} The distribution of number of objects for all candidates (thin lines), and for candidates with a mean R/B score $>$0.4 (thick lines).
{\it Bottom left:} The relation between the standard deviation ($\sigma_{\rm R/B}$) and mean ($\overline{S_{\rm R/B}}$) of the R/B scores of a candidates objects with at least 10 objects each ($C_{\rm obj}>10$).
{\it Bottom right:} The relation between a candidate's number of objects ($C_{\rm obj}$) and the mean of their R/B scores ($\overline{S_{\rm R/B}}$), for all candidates.
\label{fig:9}}
\end{figure*}

In this work, the term {\it candidate} refers to the set of associated objects at a given sky coordinate.  When the pipeline detects an object within 2\arcsec\ of a previously detected candidate, it will associate that object what that previous candidate; otherwise, it will treat the object as a new candidate.   
The two main properties of interest for candidates are the number of objects they have ($C_{\rm obj}$; i.e., the number of times they were detected in a difference image), and the mean and standard deviation in the R/B score of those objects ($\overline{S_{\rm R/B}}$ and $\sigma_{\rm R/B}$).
The top row of Figure~\ref{fig:9} shows the distributions of candidates' $\overline{S_{\rm R/B}}$ and $C_{\rm obj}$ values, and the bottom row shows the relationships between $C_{\rm obj}$, $\overline{S_{\rm R/B}}$, and $\sigma_{\rm R/B}$.

In total we identified $>600000$ candidates based on the objects detected in difference images for the COSMOS and ELAIS fields.
Many of these will be artifacts or moving objects that appear in only a few exposures (i.e., low $C_{\rm obj}$ with any $S_{\rm R/B}$), or stationary recurrent artifacts that always appear in the same place (i.e., large $C_{\rm obj}$, low $S_{\rm R/B}$) due to, e.g., detector issues or bright stars.
To identify a subset of "probably-real" candidates we establish cuts on $C_{\rm obj}$, $\overline{S_{\rm R/B}}$, as described below.

Unless an extragalactic transient or variable is rapidly changing in brightness within a couple of hours, then it should be detected in all five images of a given filter during the night (unless observing conditions were also changing within hours-long timescales, of course).
The spectral energy distributions for some transients have sharp features, such as strong hydrogen emission lines at $\lambda$6563~\AA\ for Type\,IIn supernovae or AGN, which means they might be detected in only one filter -- but these objects are also known to vary on timescales of days, and they would be detected over multiple nights.
Furthermore, a candidate would only be scientifically useful (i.e., a crude classification could be attempted) if it was detected in at least two filters, or in one filter but on at least two nights.
For these reasons, we characterize candidates as "probably-real" if they have $C_{\rm obj}\geq10$ (i.e., they were detected in at least 10 difference images, in any filter, at any time).
In the upper left panel of Figure~\ref{fig:9} we can see that the limit of $C_{\rm obj}\geq10$ rejects more candidates with low $\overline{S_{\rm R/B}}$ than with high $\overline{S_{\rm R/B}}$ (keep in mind the $y$-axis is in log-space), indicating that this cut is effective in removing artifacts.

In Section~\ref{ssec:proc_srcdet}, the minimum of the histogram of $S_{R/B}$ values for all objects at $\sim$0.6, as seen in the top row of Figure~\ref{fig:7}, suggested that $S_{R/B}\sim0.6$ could be used to identify likely-real objects.
However, in order to identify potentially-real \emph{candidates}, a lower limit on the \emph{mean} R/B score of $\overline{S_{\rm R/B}}\geq0.4$ appears to be more appropriate, for two reasons.
One, the distribution of mean R/B score values flattens out starting at $\overline{S_{\rm R/B}}\approx0.4$, as seen in the top-left panel of Figure~\ref{fig:9}.
Two, the relationship between standard deviation in R/B score and mean R/B score reaches a peak of $\sigma_{\rm R/B}\approx0.3$ at $\overline{S_{\rm R/B}}\approx0.4$, as seen in the bottom-left panel of Figure~\ref{fig:9}.
This suggests that there is more {\it certainty} of an overall low R/B score for candidates with $\overline{S_{\rm R/B}}<0.4$ -- in other words, candidates with $\overline{S_{\rm R/B}}<0.4$ are bogus with greater certainty.

Imposing the conditions that both $C_{\rm obj}\geq10$ and $\overline{S_{\rm R/B}}\geq0.4$ decreases the number of candidates for consideration from $>600000$ to 4413 -- in other words, we identify $<$1\% of the cataloged candidates as "probably real".
In the bottom-right panel of Figure~\ref{fig:9} we can see the expected trend between $C_{\rm obj}$ and $\overline{S_{\rm R/B}}$ emerge within this region of parameter space (black lines), further reinforcing these cuts as identifying "probably real" candidates.

However, we also stress that this "probably-real" sample still likely includes many artifacts (from, e.g., bright stars, as discussed for objects in Section~\ref{ssec:proc_srcdet}), and also likely excludes real astrophysical phenomena with poor R/B scores or that are very short duration (appear in $<$10 images).
This sample is being created and made available (Section~\ref{sssec:proc_canddet_github}) primarily to lower the barrier to access for anyone who is unfamiliar with how to use alerts and brokers, but who wants to explore some of the data from this DECam DDF program.

\subsubsection{Nightly-Epoch Light Curves}\label{sssec:proc_canddet_nelc}

Since creating nightly coadded images and difference images -- and generating alerts from them -- remains a stretch goal (Section~\ref{ssec:proc_egal}), we have instead combined the photometry for all detections (objects) of a candidate obtained in a given night (per filter), for the 4413 "probably-real" candidates.
Forced photometry in the difference images also remains a stretch goal (Section~\ref{sec:conc}), and so the nightly-epoch photometry is simply a mean of the difference-image \textit{detections} in a given night, and information from difference images in which the source is undetected is not yet included.

For these "nightly-epoch" light curves we calculate four parameters: the minimum magnitude (brightest observation), amplitude (difference between the minimum and maximum magnitude {\it in the difference image}), time span (days between the first and last epoch), and number of epochs of detection.
These four parameters are determined for each of the three filters, $g$, $r$, and $i$, as well as any filter.

\begin{figure*}
\includegraphics[width=17cm]{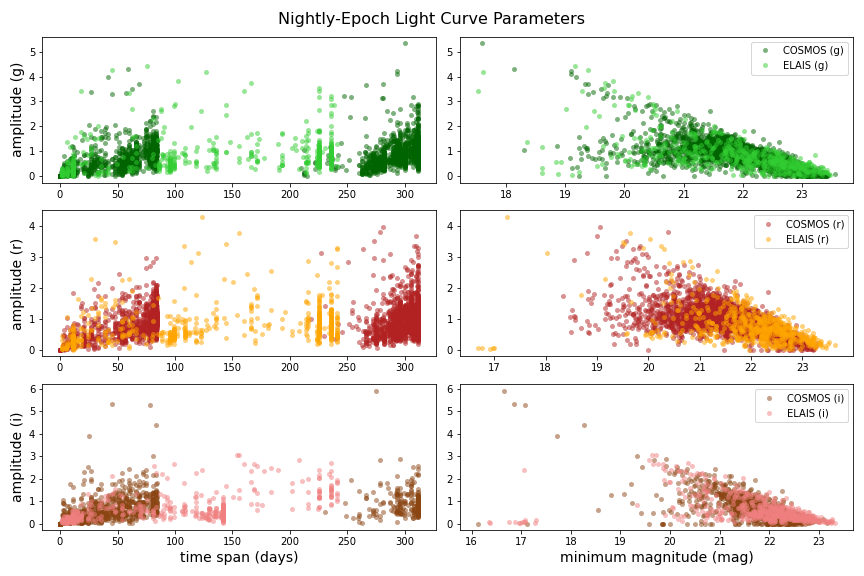}
\caption{The correlations between the variable-component's amplitude (in magnitudes), time span, and minimum magnitude as measured from the difference-image nightly-epoch light curves of the "probably real" candidates in the extragalactic fields, COSMOS and ELAIS, for each filter ($g$, top; $r$, middle; and $i$, bottom).
The full light curves and the derived parameters for all of these candidates are available online as described in Section~\ref{sssec:proc_canddet_github}.
\label{fig:10}}
\end{figure*}

In Figure~\ref{fig:10} we show the correlations between these 4413 candidates' nightly-epoch light curves' amplitude, time span, and minimum magnitude for each filter.
In the left-column panels we can see that many of the "probably real" candidates have a long duration (e.g., AGN, variable stars), and that the time spans cluster around the season lengths for the extragalactic fields.
For example, the 2021A season for COSMOS was $\sim$90 days long (March to June), and then the next observations were in January 2022, causing clumps of COSMOS data points at 90 and 310 days in the lower-left panel.
In the right-column panels of Figure~\ref{fig:10}, as expected we see that most of these candidates' light curves' brightest magnitude were faint ($>$22 mag), and that most candidates have a low amplitude ($<$0.5 mag), but there is clearly more to explore in this data set.

In Section~\ref{sec:sci} we demonstrate how simple cuts on these nightly-epoch light curve parameters make good starting points for identifying samples of transients and variables for further analysis, and how some preliminary classifications and scientific investigations that can be done with these parameters.

\subsubsection{Public Data Products for 4413 Candidate Light Curves}\label{sssec:proc_canddet_github}

Files containing data for the 4413 "probably-real" candidates that have been identified, as described above, in difference-images of the extragalactic fields COSMOS and ELAIS during 2021 are available on {\tt GitHub}\footnote{See Version 2 of the {\tt decam\_ddf\_tools} repository at \url{www.github.com/MelissaGraham/decam_ddf_tools/tree/v2.0}}.
Keeping in mind that the term "object" means a difference-image detection, the data for the 4413 candidates is available in five files which include:
\begin{itemize}
    \item Exposure metadata such as date, seeing, limiting magnitude, number of objects.
    \item Candidate metadata such as coordinates, number of associated objects and their mean R/B score.
    \item All associated objects' date, filter, magnitude, and R/B score ($>300000$ objects, total). These are the single-image light curves.
    \item Candidate nightly-epoch light curves (combined difference-image detections).
    \item Summary parameters for the nightly-epoch light curves (amplitude, duration, etc.).
\end{itemize}

\noindent
A {\tt Jupyter Notebook} tutorial demonstrating how to access and plot the data in those five files is also provided\footnote{This notebook is named {\tt 01\_demo\_candidates.ipynb} in Version 2 of the {\tt decam\_def\_tools} repository.}.

The data (the images and the photometry released via {\tt GitHub}) from this DDF program are all completely public, and everyone should feel free to pursue science with this dataset even if they are not a co-investigator or co-author, and even if their science overlaps with some of the examples provided in Section~\ref{sec:sci}.
All users should keep in mind the caveat that the photometry being made available in these files is based on detections in difference images only, and that measurements in the direct or template images are not available (i.e., no "total" fluxes).
Difference-image photometry is most appropriate for use with transients, which do not appear in the template image, whereas, e.g., variable star or AGN studies typically generate light curves from the direct-image flux.

\section{Preliminary DECam DDF Science}\label{sec:sci}

The main purpose of this paper is to present the DDF survey data for 2021, but to further illustrate the scientific potential of this dataset we provide preliminary science results in our three main science areas: the Solar System (Section~\ref{ssec:sci_ss}); Galactic stellar variables and transients (Sections \ref{ssec:sci_gal_flares} through \ref{ssec:sci_galegal_fast}); and extragalactic transients and variables (Sections \ref{ssec:sci_galegal_fast} through \ref{ssec:sci_egal_sl}).
Some of these preliminary results use the same pipeline data products (candidates and objects) presented above, some use the alerts, and some are based on independent analyses of the images.

\subsection{Solar System Science: the discovery and characterization of Main Belt Asteroids and Trans-Neptunian objects}\label{ssec:sci_ss}

\textit{Co-authors: Stetzler, Smotherman, Heinz, and Juric.}

\smallskip \noindent

Sources in difference images from the COSMOS-1, COSMOS-2, COSMOS-3, and DECaPS-East fields are being analyzed to discover and characterize main belt asteroids and their colors. 
Instead of using the data products described in Section~\ref{sec:proc}, difference images and source catalogs of the survey data for these fields were produced using the LSST Science Pipelines\footnote{\url{https://pipelines.lsst.io}} \citep{2017ASPC..512..279J}.
These fields are chosen based on their low ecliptic latitudes.
The DECaPS-East field is particularly interesting as it images very dense fields, producing large source catalogs that are expected to strain asteroid detection algorithms.
Additionally, the $z$-band imaging of the DECaPS-East field provides magnitude measurements and colors that can potentially differentiate dynamical classes of main belt asteroids.

The difference images from these fields will also be searched for slow-moving Trans-Neptunian objects (TNOs) using the digital tracking code \texttt{KBMOD} \citep{Whidden_2019,Smotherman_2021}.
\texttt{KBMOD} is a GPU-accelerated software package that searches a large grid of possible moving object trajectories.
In previous works, \texttt{KBMOD} modelled moving object candidate trajectories as lines in topocentric space.
As such, it has been constrained to survey time-baselines of hours-to-days.
Recently, \texttt{KBMOD} has been extended to model candidate trajectories as lines in barycentric space using the cartesian coordinates of the observatory, Earth, and Solar System barycenter.
This improvement should allow \texttt{KBMOD} to search over a time baseline of several months and will be necessary for applying digital tracking to LSST.
As such, we will apply this technique first to the survey described in this paper.

\subsection{Galactic Science: constraining stellar flare temperatures}\label{ssec:sci_gal_flares}

\textit{Co-authors: Clarke, Bianco, and Davenport}.

\smallskip \noindent
Stellar flares have short timescales (minutes to hours) and preferentially occur on low-mass stars \citep[e.g.,][]{2018ApJ...853...59C,2019ApJ...871..241D}.
Characterizing the temperatures and energies of stellar flares are the next astrophysical frontier in flare studies with time-domain surveys.
Deep drilling multi-band surveys, like this one with DECam and the future DDF observations done as part of the LSST with Rubin Observatory, provide both the time resolution on minutes-long timescales to identify flares, and the requisite colors to use as flare temperature indicators.

To start, we have identified candidates in the COSMOS or ELAIS fields with two or more objects -- \textit{all in the same night} -- at least one of which is in $g$-band (because of the expected flare temperature), and a mean real/bogus score of $\overline{S_{\rm R/B}} \geq 0.6$.
We are not yet using the Galactic fields because their R/B scores are not yet as well-characterized as the extragalactic fields' are.
The 1273 candidates which meet these conditions are currently being examined to determine whether they might be stellar flares.

In the future, we also plan to use astrometric data and the differential chromatic refraction (DCR) effect for observations at moderate to high airmass as a way of confirming single-image observations are likely to be stellar flares, and to potentially constrain their temperature. 
Since the DCR effect imparts a known wavelength and airmass dependent shift to a source along the parallactic angle, we are able to reconstruct weak spectral information for an event based on the information available within the alerts (e.g. MJD, RA, Dec).
For this reason, the 5--20\% of observations obtained at airmass $>$1.4 (Figure~\ref{fig:2}) might be particularly useful for the study of stellar flares, even though the survey strategy aims primarily for low-airmass observations.
At the time of publication, stellar flare work with this DECam DDF program was ongoing, and planned to be presented in a future paper.

\subsection{Galactic Science: characterizing variables in the DECaPS-East field}\label{ssec:sci_gal_vars}

\textit{Co-authors: Patel and Soraisam}.

\begin{figure*}
\includegraphics[width=9.03cm]{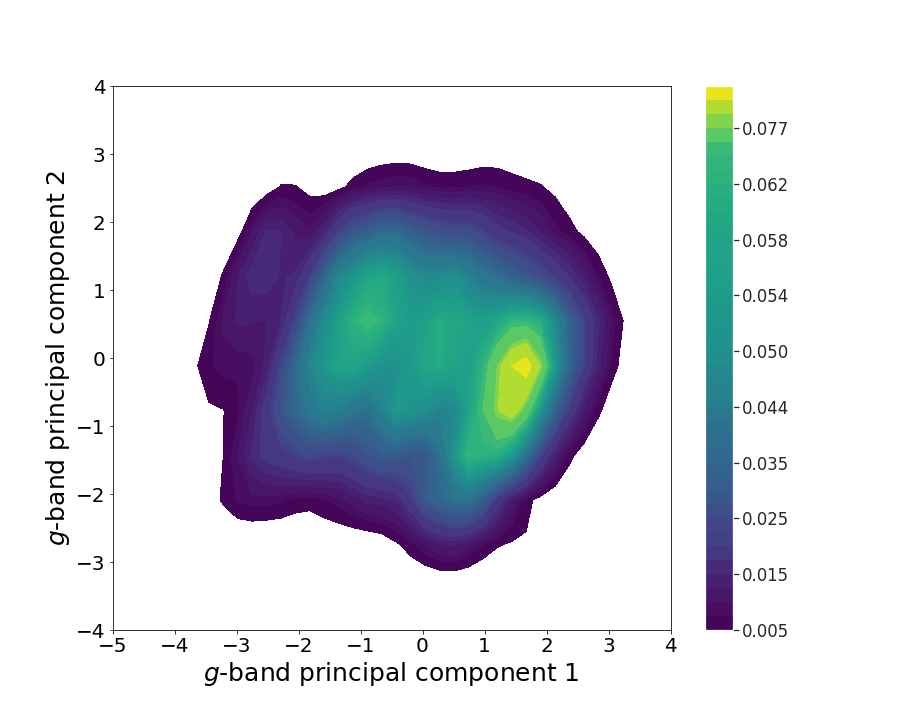}
\includegraphics[width=7.5cm]{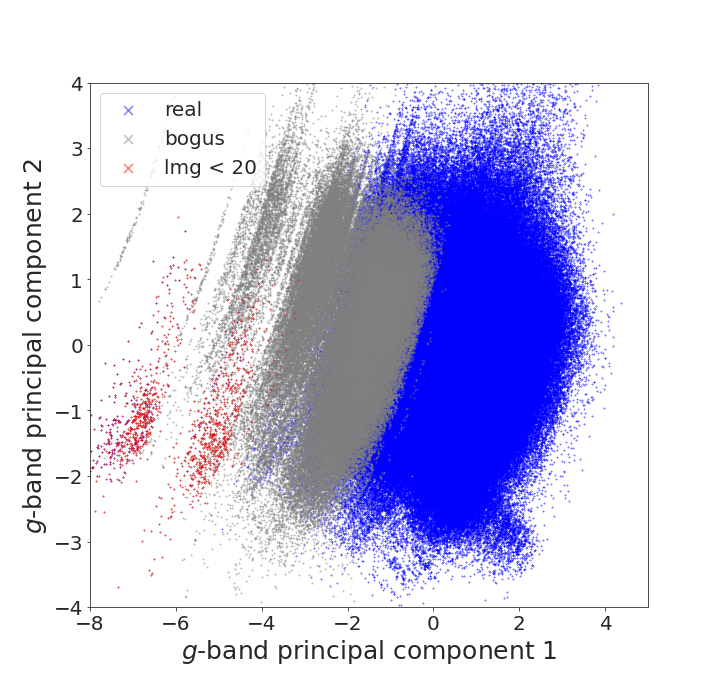}\\
\includegraphics[trim={0 0 0 1cm},clip,width=9.03cm]{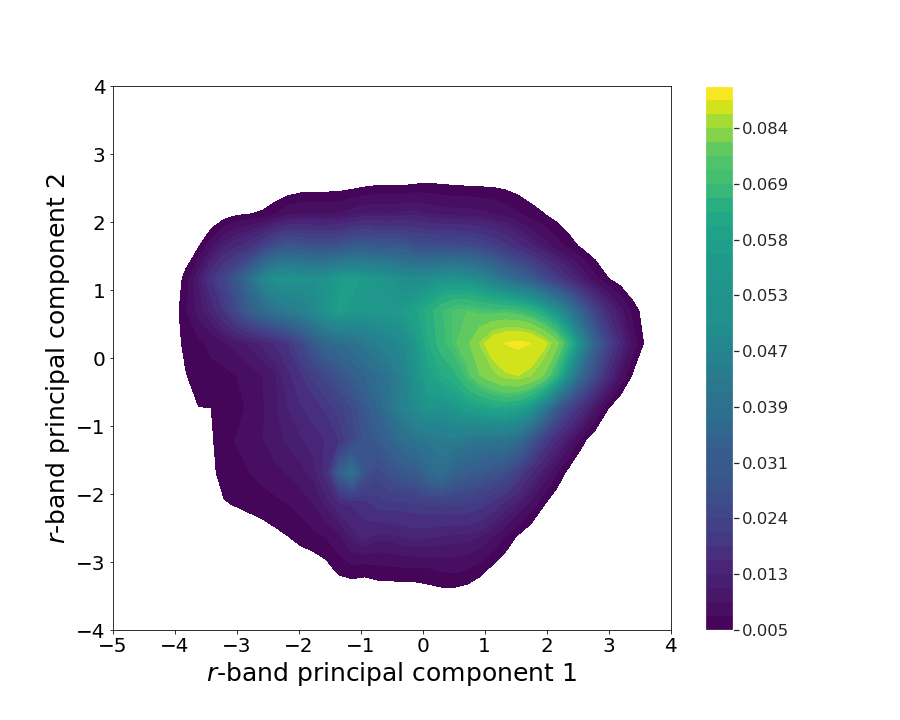}
\includegraphics[trim={0 0 0 2cm},clip,width=7.5cm]{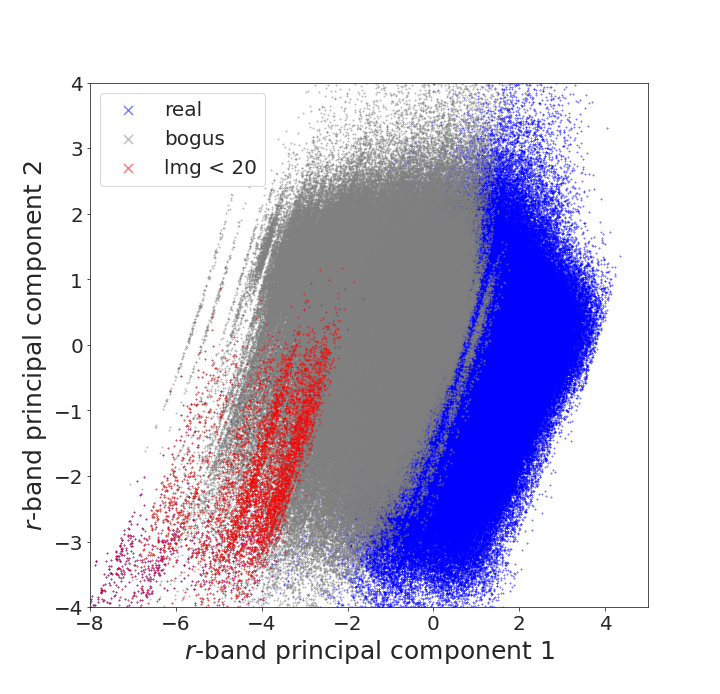}\\
\includegraphics[trim={0 0 0 1cm},clip,width=9.03cm]{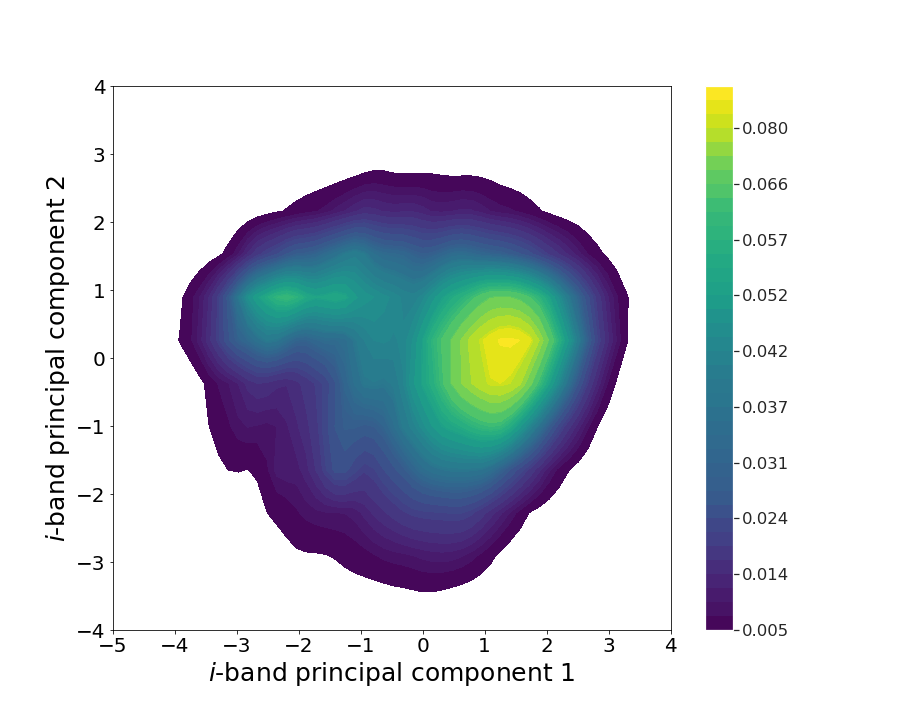}
\includegraphics[trim={0 0 0 2cm},clip,width=7.5cm]{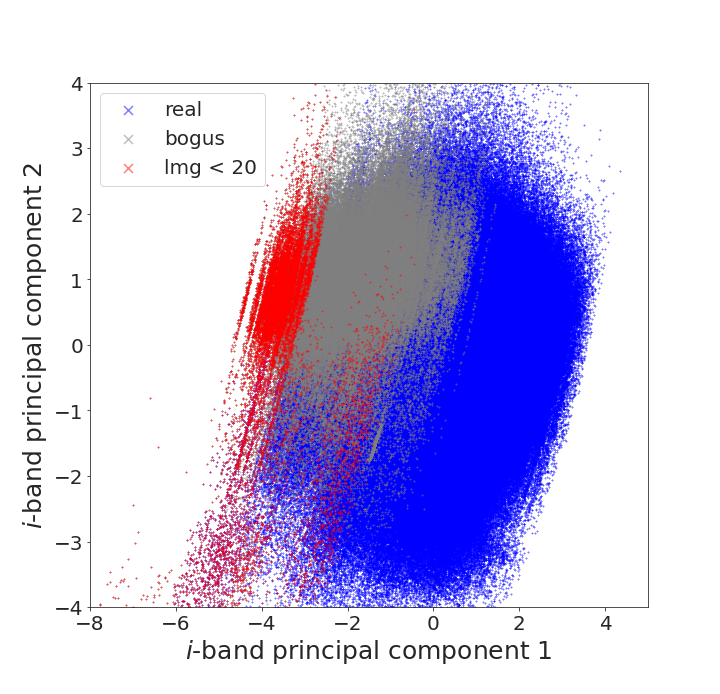}   
\caption{Principal component analysis of the Galactic alerts data.
The $g$, $r$, and $i$-band results are shown in the top, middle and bottom rows, respectively.
Plots in the left panels represent the density of alerts, while the right panels are scatter plots with the predictions from the Gaussian Mixture model (GMM) clustering shown in grey and blue.
Visually vetted bogus alerts (highlighted by the red points) predominantly overlap with the grey points. We therefore label the grey cluster as bogus and the blue cluster as real. In the legend, \textit{lmg} is the limiting magnitude for the alerts data.}
\label{fig:11}
\end{figure*}

\begin{figure}
\centering
\includegraphics[width=8.5cm]{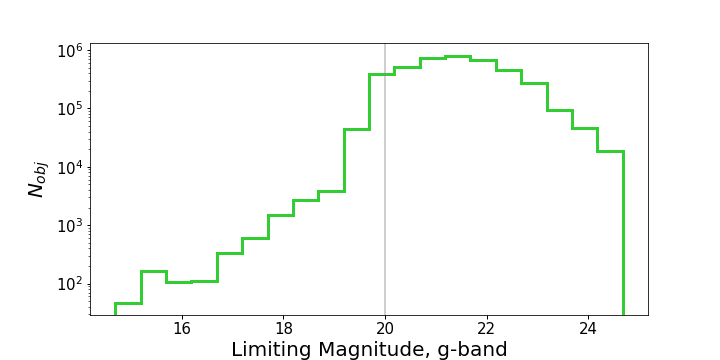}
\includegraphics[width=7cm]{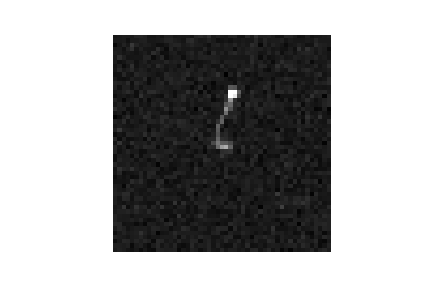}
\caption{\textit{Top:} Distribution of limiting magnitude for alerts in the DECaPS-East region in $g$-band. We obtain a similar distribution for other passbands. Based on visual inspection of difference image cutouts, we find that the alerts with limiting magnitude brighter than $\approx 20$ (marked by the grey line) are largely bogus. \textit{Bottom:} An example difference image cutout of a bogus alert. 
}
\label{fig:11.5}
\end{figure}

\smallskip \noindent
For this preliminarry study we have chosen to focus on the data from just the first semester, 2021A, during which only the DECaPS-East Galactic field was observed (Figure~\ref{fig:1}).
We obtain a list of 190,244 "good" candidate variables from the DECaPS-East field via the procedure explained below.

As described in Section~\ref{ssec:proc_RB}, the R/B classifier for detections (alerts) in the Galactic fields had not yet been fully trained at the time of publication, and the R/B classifier trained for the extragalactic fields performs poorly for detections in the Galactic fields.
Instead, we have performed a coarse unsupervised classification of these events to remove the bogus detections, and we plan to train a deep learning-based R/B classifier specifically for the Galactic alerts, which we expect to outperform this initial classification.

After exploring the available features of the alert data, we selected seeing, sky background, magnitude, magnitude error, and limiting magnitude to perform the R/B classification.
We then apply Principal Component Analysis (PCA) using the {\tt decomposition} module in the {\tt scikit-learn} package to reduce the dimensionality of our feature space.
For each passband, we obtain a total explained variance of roughly 80\% based on the first two principal components.
Also, we find that sky background, limiting magnitude and magnitude error of the alert are the most important features in the PCA for each passband.

Figure~\ref{fig:11} shows the PCA results for each passband for the DECaPS-East alerts.
Finally, we run a Gaussian Mixture model (GMM) with two Gaussian components over these results to cluster the alerts; the predictions are color-coded by blue and grey in Fig.~\ref{fig:11}.
During the exploratory phase of our analysis, we found that alerts with limiting magnitude brighter than 20 were heavily contaminated by cosmic rays and satellite trails after examining their difference image cutouts (see Fig.~\ref{fig:11.5}).  
To assign {\it real} and {\it bogus} labels to these clusters, we determine which cluster contains the larger percentage of these known bogus detections and label it as bogus and the remaining cluster as real. After aggregating the real alerts into unique objects (i.e., constructing the light curves), we deploy a second level of filtering, removing objects that have one alert since those are likely cosmic ray hits.
The resulting sample of objects are our "good" candidates.

\begin{figure*}
\centering
\includegraphics[width=17cm]{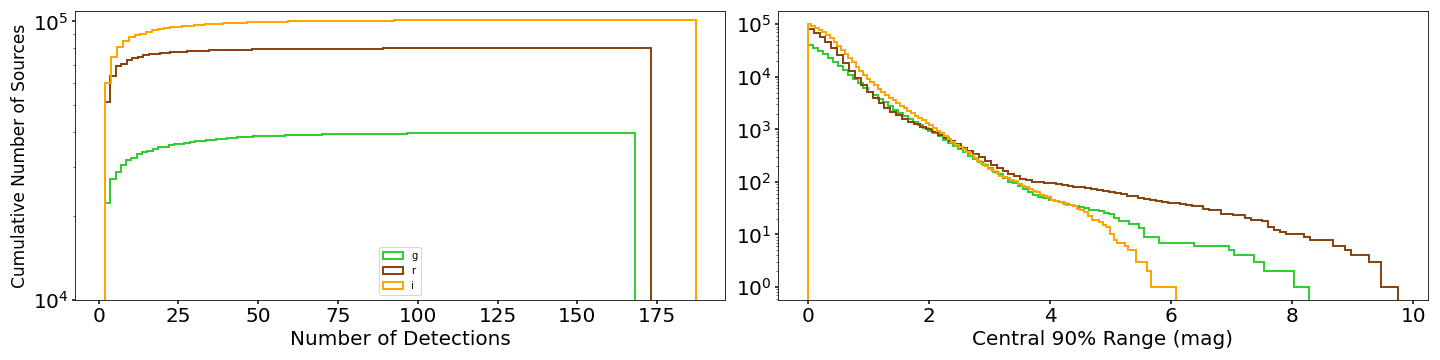}
\caption{\textit{Left}: Cumulative distribution of the number of observations (less than or equal to given \textit{number of detections}) obtained for each variable star.
\textit{Right}: Cumulative distribution of the central 90\% ranges of the magnitude distributions for our list of "good" candidates.
The distribution shows the number of sources greater than or equal to a given central 90\% range mag.}
\label{fig:12}
\end{figure*}

The cumulative distribution of the number of detections recorded for each candidate is shown in Figure~\ref{fig:12}.
We observe that there are typically tens of detections in the light curve of a candidate for all bands, which indicates that our sample contains largely persistent variable stars. 
We also show the cumulative distribution of variability amplitudes estimated using the central 90\% range of the magnitude distribution for each candidate (see \citealt{2020ApJ...892..112S}) in Fig.~\ref{fig:12} (right panel). 

\begin{figure}
\centering
\includegraphics[width=8cm]{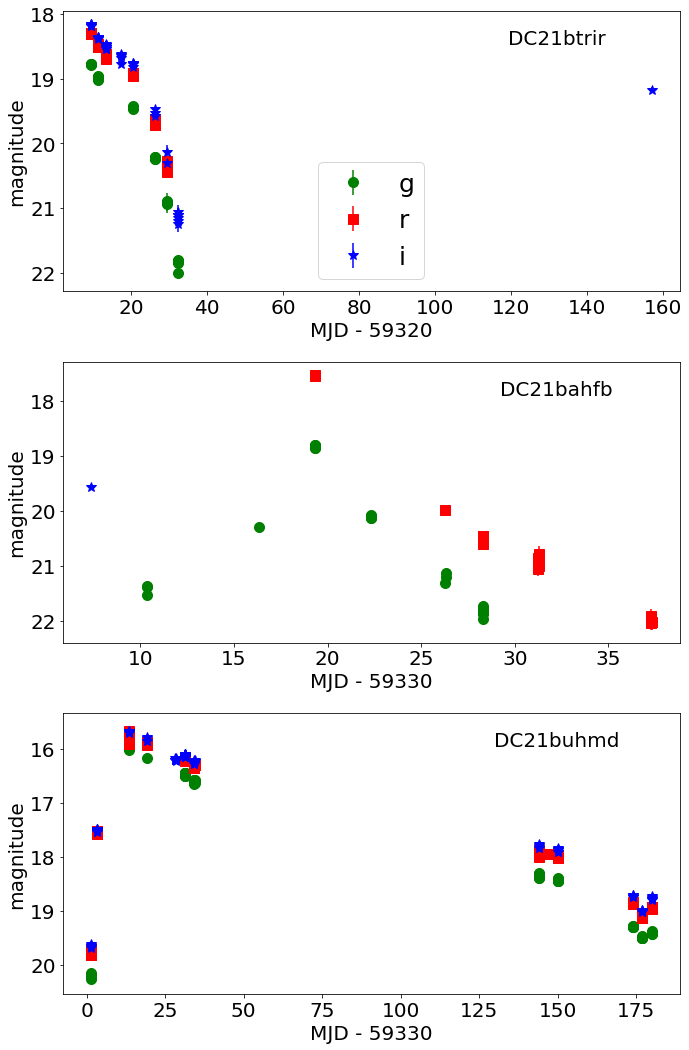}
\caption{Multi-wavelength light curves for three selected anomalous candidates.
The candidate identifiers are provided in the upper or lower right corner of each panel and the second panel in the top row provides the legend indicating the filter (green circles for $g$, red squares for $r$ , and blue stars for $i$) for the photometry points.}
\label{fig:13}
\end{figure}

We cross-matched our good candidates with the Gaia Data Release 3 (DR3) sample of variable stars \citep{2022arXiv220606416E} and found 5667 matches.
The classifications of these variable stars are also included in \citet{2022arXiv220606416E}.
To avoid misidentifying our candidates when cross-matching crowded fields such as the Galactic Bulge, we choose a conservative radius of $1\arcsec$ for our list of candidates.
Of the matches, we find 3143 are long period variables (LPV), 1671 are eclipsing binaries (ECL), 511 are RR Lyrae 
(RRLyr), 12 are short timescale variables, and 4 are Cepheids (CEP).
We also cross-matched our good candidates with the variable star catalogs from the Optical Gravitational Lensing Experiment (OGLE) in the Galactic bulge \citep{2008AcA....58...69U,2015AcA....65....1U, 2018AcA....68..315U, 2014AcA....64..177S, 2019AcA....69..321S, 2022ApJS..260...46I}.
We found 4480 matches after applying a conservative search radius of $1\arcsec$ for our candidates.
We found that 3835 are LPVs, 639 are RRLyrs and 5 are CEPs.
In the future, we can use these data sets as the training samples to perform ML classification of the remaining variables. 

We further use the algorithm developed by \cite{2020ApJ...892..112S} to select anomalous sources from our sample of good candidates, making use of the multi-band time series data from DECam.
Their algorithm measures a likelihood score of consistency of the features of a given light curve with those of the parent sample; light curves with the lowest scores are flagged as anomalies (see \citealt{2020ApJ...892..112S} for details). 

This anomaly detection algorithm only uses neighboring pairs of observations.
We add a visual inspection step of the identified outliers, which allows us to include information on the overall shape of the light curve in a qualitative way and exclude sources that, albeit not being typical, are identifiable as members of a class of objects that in itself is not rare.
For example, LPVs with extreme amplitudes may be flagged as outliers but are not of particular interest for our study.

After visual inspection of the light curves for a few hundred of the most anomalous sources, we select three interesting candidates, whose light curves are shown in Fig.~\ref{fig:13}. 
We find that DC21btrir declines by 3 mag in roughly 30 days in the $g$ and $i$-bands.
It was first classified as a CV by \cite{2001PASP..113..764D}.
DC21bahfb has a symmetric outburst-like profile, characteristic of microlensing events lasting about tens of days.
It is likely a new microlensing event.
DC21buhmd is the optical counterpart of the low-mass X-ray binary MAXI J1803-298 first detected in \cite{2021ATel14587....1S} and followed up in the optical band by \cite{2021ATel14706....1S}.
The multi-band time series data of its optical outburst have been captured by the DDF survey.

\subsection{Galactic Science: multi-band light curve templates }\label{ssec:sci_gal_var-temps}

\textit{Co-authors: Catelan, Rodríguez-Segovia,  and Baeza-Villagra}.

As pointed out in the previous subsection, DDF data can be used to detect and characterize a large number of variable stars of different types.
Due to its long (and expanding) time coverage (Fig.~\ref{fig:1}) plus large (and increasing) number of observations (Table~\ref{tab:2}), depending on their periods, complete phase coverage can be achieved for at least some periodic variable stars.
Our multi-band approach thus provides the opportunity of obtaining the template light curves that will be required to properly inform the next generation of {\em multi-band classifiers} that will be required to fully realize Rubin/LSST's potential. 

PSF photometry of the DECaPS-East field DDF images obtained during semesters 2021A and 2021B\footnote{We included the short-exposure images obtained during bright time in 2021, which are discussed in Section~\ref{sssec:survey_char_bright}.} was carried out using an implementation of the {\tt photpipe} pipeline modified for DECam images.
{\tt photpipe} is a robust pipeline used by several time-domain surveys \citep[e.g., SuperMACHO, ESSENCE, Pan-STARRS1; see][]{Rest2005,Rest2014}, designed to perform single-epoch image processing including image calibration (e.g., bias subtraction, cross-talk corrections, flat-fielding), astrometric calibration, warping and image coaddition \citep[{\tt SWARP};][]{2002ASPC..281..228B}, and photometric calibration.
Additionally, {\tt photpipe} performs difference imaging using {\tt HOTPANTS} \citep{Alard2000, 2015ascl.soft04004B} to compute a spatially varying convolution kernel, followed by photometry on the difference images using an implementation of {\tt DoPHOT} \citep{DOPHOT1, DOPHOT2} PSF photometry on difference images \citep{Rest2005}.
In this work, however, difference imaging has not yet been used. 

\begin{figure*}
\centering
\includegraphics[width=17cm]{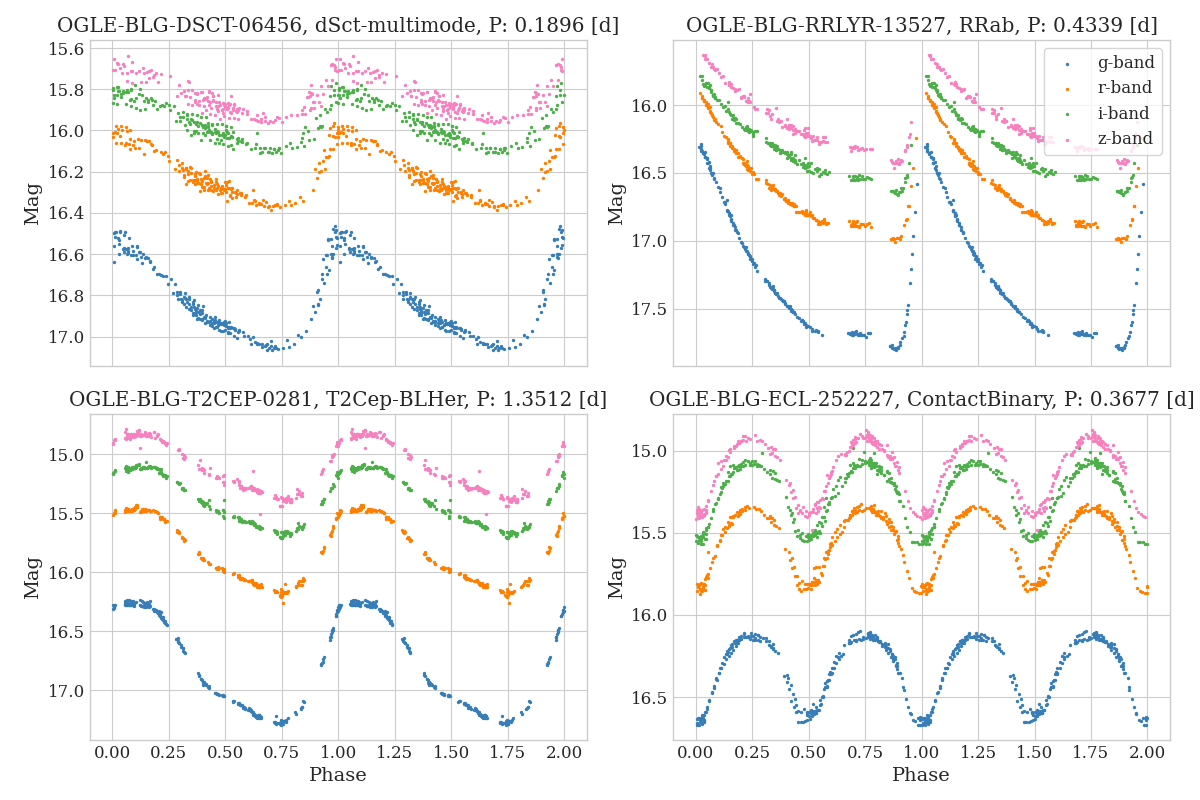}
\caption{Multi-band light curves, obtained from DDF data for the DECaPS-East field.
Observations in the $g$, $r$, $i$, and $z$ bands are shown in blue, orange, green, and pink, respectively.
From upper left to lower right, one finds a multi-periodic $\delta$~Scuti with a dominant period of 0.1896~d, an ab-type RR Lyrae with a period of 0.4339~d, a short-period type II Cepheid (or BL Her star) with a period of 1.3512~d, and a contact binary with a period of 0.3677~d, respectively.
Each star's OGLE ID is shown at the top of its corresponding panel, along with its variability class and period. }
\label{fig:14}
\end{figure*}

Examples of the phase-folded light curves that were obtained following this approach are provided in Figure~\ref{fig:14}, where only measurements having {\tt dotype = 1} (corresponding to stellar sources) are included.
In this plot are shown our $griz$ light curves for each of the following variables, which had previously been studied in different phases of the OGLE project \citep{OGLE-RRLYR,OGLE-T2Cs,OGLE-EBs,OGLE-DSCT}: OGLE-BLG-DSCT-06456, OGLE-BLG-RRLYR-13527, OGLE-BLG-T2CEP-0281, and OGLE-BLG-ECL-252227. The adopted periods are the same as reported by the OGLE team.
We are currently extending this work to other previously known variables in the DECaPS-East field.
A search for, and classification of, other unknown variables in the same field is also underway.
In the future, data from other DECam programs covering the same field (e.g., program 2021A-0921, P.I. M. Catelan) will be used to extend the phase coverage, and the corresponding photometry will be made public as well (Catelan et al. 2022, in preparation).

\subsection{Galactic and Extragalactic Science: detecting first-night fast evolvers}\label{ssec:sci_galegal_fast}

\textit{Co-authors: Kennedy and Graham}.

\smallskip \noindent
Being able to identify -- and spectroscopically follow-up -- a new fast-evolving transient or variable as early as possible has a wide variety of astrophysical use-cases.
As this DDF program does at least 15 exposures on the same area of sky during the night, it should be well-suited to finding new fast-evolving events.

We start with the set of all COSMOS- and ELAIS-field candidates (not just the "probably-real" candidates described in Section~\ref{ssec:proc_canddet}), and then select only those which had $\geq$5 objects (detections) with R/B scores $>$0.4 in any filter during first night they were detected, and for which the median magnitude error of these objects was $\leq$0.03 mag.
In other words, we first limit to candidates that have high-quality photometry (i.e., are bright), are well-sampled, and are likely-real (not bogus) detections on their first night of detection.
We then fit a line to those objects using {\tt numpy.polyfit}, and flagged each candidate for which we detected a rise at the 2-sigma level (using the returned covariance matrix) as a potential "fast riser."
Under these conditions we identify 24 potential fast-risers.

Eleven of these candidates were detected in only one night, two of which (DC21efoi and DC21fbia) were readily identified as asteroids using the Minor Planet Center's MPChecker\footnote{\url{www.minorplanetcenter.net/cgi-bin/checkmp.cgi}}.
These two candidates, plus two others (DC21kqjn and DC21lktc) of the eleven, met the constraints used to identify "probably-real" candidates in Section~\ref{ssec:proc_canddet}; the remaining seven did not satisfy the requirement of $C_{\rm obj}\geq10$.
As we are interested in how well we can identify multi-night variables and transients that rise quickly in their first night, these one-night-only candidates are not investigated further in this work.

For the thirteen identified potential fast-riser candidates with detections over more than one night, we found that the intra-night best-fit line always overpredicted the brightness of the next observation by a large margin.
This suggests that these candidates are not explosive transients (i.e., supernovae) but are more likely to be compact objects or stars, for which short timescale variability is not always representative of a days-long trend.
We acknowledge that the constraint that we must place on a candidate having "high-quality photometry" with low magnitude errors (so that we can reliably identify a positive slope with only 5 data points) means that we are also limiting this test to bright variable objects, which is a bias against extragalactic transients.
We will explore more specific methods for detecting \emph{faint} fast-risers in future work.

Seven of the thirteen multi-night candidates flagged by these cuts also met our "probably-real" conditions and have nightly-epoch light curves available online (see Section~\ref{ssec:proc_canddet}): DC21kkqh, DC21kldj, DC21kluc, DC21koer, DC21kptk, DC21krys, DC21kvqx.
The remaining six did not satisfy the constraint on the mean R/B score, $\overline{S_{\rm R/B}}\geq0.4$, and are not discussed further in this work.
Of the seven "probably-real" candidates, only DC21kkqh is listed in the CDS\footnote{The Strasbourg astronomical Data Center.} SIMBAD\footnote{Set of Identifications, Measurements, and Bibliography for Astronomical Data. \url{https://simbad.u-strasbg.fr/simbad}} catalog as a QSO\footnote{Quasi-Stellar Object}, but all seven show point sources in their reference images as additional indication that all seven are stellar variables.

\begin{figure}
\includegraphics[width=8.5cm]{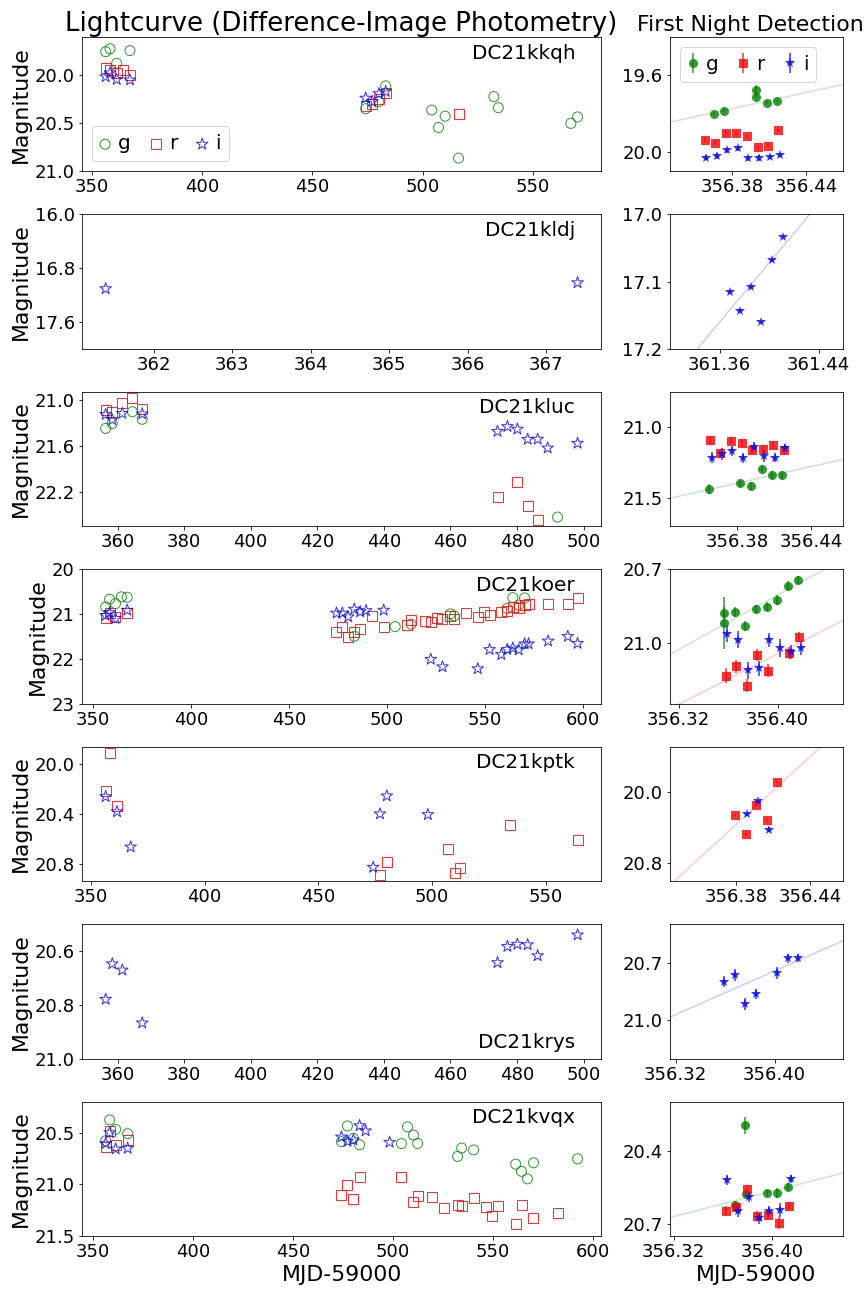}
\caption{The seven "probably-real" first-night fast-rising candidates (one per row).
The left column of panels shows the nightly-epoch photometry (error bars omitted for clarity), and the right column shows the individual-image photometry during their first night of detection.
For DC21koer (fourth row), two trend lines are displayed in their first-night photometry panel because this candidate was flagged as a fast-riser in both $g$- and $r$-bands.
Candidate identifiers for each row are displayed on the right side in the left column of panels.
\label{fig:15}}
\end{figure}

To illustrate the first-night "fast-rising" detections in context with the full photometric data set for these seven "probably-real" candidates, we display their light curves in Figure~\ref{fig:15}.

\subsection{Extragalactic Science: potential Type Ia supernovae}\label{ssec:sci_egal_sneia}

\textit{Co-authors: Graham and Kennedy}.

\begin{figure}
\includegraphics[width=8.5cm]{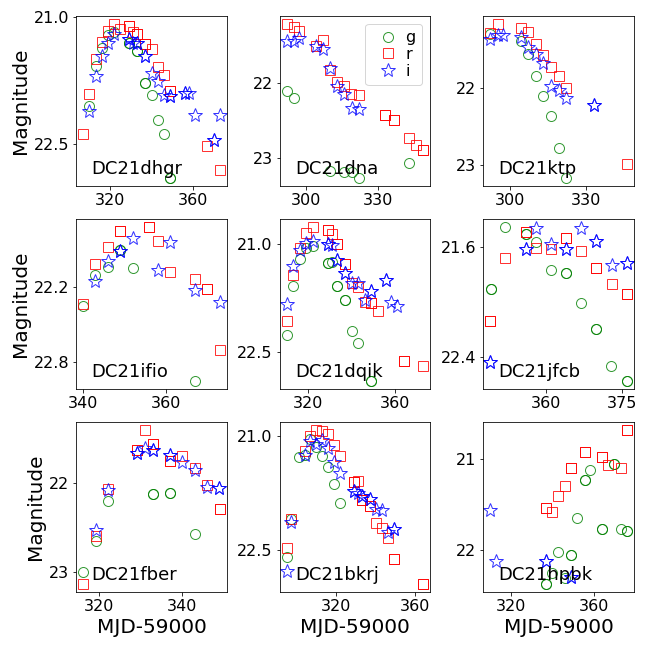}
\caption{Nine of the twenty-two potential SN\,Ia (randomly chosen) that were identified among the "probably-real" candidates with a very simple series of cuts on the nightly-epoch light curve parameters.
The candidate identifiers are provided in the lower left corner of each panel, and the second panel in the top row provides the legend for the photometry points in each filter (green circles for $g$, red squares for $r$, and blue stars for $i$).
Magnitude error bars are omitted for clarity.
\label{fig:16}}
\end{figure}

\smallskip \noindent 
To obtain a list of \textit{potential} Type Ia supernovae (SNe\,Ia; the explosions of carbon-oxygen white dwarf stars) as a starting point for more specific light curve fitters, we make use of the fact that SN\,Ia light curves are fairly homogeneous and that there is a correlation between light curve time span and amplitude (i.e., lower-redshift SNe\,Ia are brighter than the survey's limiting magnitude for a longer time).
We consider a candidate a "potential SN\,Ia" if it has an amplitude of $>0.5$ mag and a time span of $>10$ days in each filter and also has a sufficiently large amplitude for its time span (as defined by "normal" SN\,Ia light curves from \citealt{2002PASP..114..803N}).
With this definition of "potential SN\,Ia", we obtain 22 candidates for further consideration from the 4413 "probably-real" candidates described in Section~\ref{ssec:proc_canddet}.
The nightly-epoch light curves for nine of these candidates are shown as a demonstration in Figure~\ref{fig:16}.

This is just an example of what simple cuts on the light curve parameters can provide, and a demonstration that there are plenty of SN-like candidates in the database with well-sampled light curves -- this is by no means a confirmation that these objects \textit{are} SNe\,Ia.
The next steps of performing light curve template fits and/or machine-learning classification of these transients is left to future work.

\subsection{Extragalactic Science: detecting variability in known Active Galactic Nuclei (AGN)}\label{ssec:sci_egal_agn}

\textit{Co-authors: Kennedy and Graham}.

\smallskip \noindent
Variability across all bands is a key characteristic of active galactic nuclei (AGN) (see, e.g., \citet{2017A&ARv..25....2P}).
To quantify the portion of AGN our survey can identify as being variable, we matched our 4413 likely-real candidates (see Section \ref{ssec:proc_canddet}) to catalogs of known AGN in our extragalactic fields\footnote{Data from all catalogs discussed here were retrieved using SIMBAD \citep{2000A&AS..143....9W}, and nearest-neighbor matching was performed with a maximum separation of 2 arcseconds}.
The COSMOS field in particular has a long history of AGN studies \citep[e.g.,][just to name a few]{2019A&A...627A..33D,2015A&A...578A.120L,2012ApJ...753...30S}, and the DDF survey strategy was set partially to share data with a DECAT program focused on long-term AGN monitoring program for COSMOS (Section~\ref{ssec:survey_fieldsel}).

Our COSMOS fields have complete overlap with the COSMOS portion of the Advanced Camera for Surveys General Catalog (ACS-GC, \citet{2012ApJS..200....9G}) and the COSMOS2015 galaxies catalog \citep{2016ApJS..224...24L}.
We anticipate at least 10$\%$ of the galaxies in these catalogs to fall within the DECam chip gaps (we do not execute a dither pattern).
We detected variability in 128/1349 (9.5$\%$) of the ACS-GC AGN sample and 164/2970 (5.5$\%$) of the COSMOS2015 AGN sample.
When we break these numbers down by AGN sub-classifications, we find that we detect approximately one third of their QSO samples as variable (72/218 and 83/267, respectively), and only slightly lower percentages for Seyfert 1 galaxies (5/17 and 6/22, respectively).
We did not detect variability in any of the galaxies classified as LINERs or Seyfert 2s in these catalogs.
Our ELAIS pointings have complete overlap with the field targeted by the ESO-Spitzer Imaging extragalactic survey (ESIS, \citet{2006A&A...451..881B}).
We detect 31/346 ($9\%$) of their AGN as optically variable, and 3/24 ($12.5\%$) of their QSOs.

More detailed work with the AGN in our DDFs, such as identifying previously unknown AGN via optical variability, or analyzing their short- and long-timescale light curves, is left to future work.

\subsection{Extragalactic Science: finding gravitationally lensed supernovae}\label{ssec:sci_egal_sl}

\textit{Co-authors: Magee and Collett}.

\smallskip \noindent
The COSMOS field is home to a number of confirmed or candidate gravitational lensing systems.
Through our continuous monitoring of this field, we aim to detect any background supernovae that are gravitationally lensed by these systems.
We plan to make use of nightly and weekly image stacks to reach deeper magnitudes (i.e., $>$25 mag) -- but as described in Section~\ref{ssec:proc_egal}, the processing pipeline for these intermediate-timescale stacks was not yet running.
Foreground supernovae will be removed by comparing against templates at the apparent host-galaxy redshift.

\section{Conclusions}\label{sec:conc}

In this paper we have presented the survey strategy and processing pipieline, and characterized the images and sources detected, for the "Deep Drilling in the Time Domain with DECam" survey in the 2021-A and -B semesters.
We have shown how observing conditions and image quality affect the number and brightness of difference-images sources that we can detect, and how various candidate parameters (R/B, number of detections) can be combined to identify "probably-real" time-variable astrophysical sources as a starting point for further analysis.
We've also provided a few examples of the ongoing science investigations being done with this program's data, which span a wide range of fields from the Solar System, to Galactic stellar science, and out to extragalactic objects.

Technical aspects to the processing that are currently under development include:
\begin{itemize}
\item \textbf{Galactic Real/Bogus:} As mentioned at various points in this paper, the R/B score is currently being retrained for the Galactic fields.
\item \textbf{Extragalactic Real/Bogus:} This work has identified a need for further characterization of the R/B scores as a function of image quality and association with persistent or systematic sources. 
\item \textbf{Nightly Stacks:} Running a nightly pipeline that stacks all the images in a given filter for that night, and then does difference imaging and source detection on the nightly stacks. This will better reveal the faint, long-duration transients in the deep fields.
\item \textbf{Cross-Matching:} Building deep catalogs of static sky sources from our first year of imaging, and cross-matching newly detected difference-image sources with them (e.g., to include DDF-derived host galaxy data in alerts).
\item \textbf{Forced Photometry:} Generating forced photometry light curves using the known locations of transients and variables in the difference-images, to push to fainter magnitudes.
\item \textbf{Broker Filters:} Developing and installing alert filters with one or more brokers, so that the public can more easily interact with the DDF alerts.
\end{itemize}

Our aim is to continue to distribute alerts in real time, and to release additional derived data products at longer latency.
The goal is for the community to use for DDF-related science now, and also to inform their preparations for future science with the LSST DDFs from the Rubin Observatory.

\section*{Acknowledgements}

This DECam program for Deep Drilling Fields is a founding member of the DECam Alliance for Transients (DECAT), a logistical solution for a heterogenous group of programs all doing time-domain astronomy on a classically-scheduled telescope.
Within DECAT, multiple DECam programs request that their awarded time be co-scheduled on the Blanco 4m telescope, and then the PIs work together to ensure the targets for all programs are optimally observed, and all program participants share in the observing responsibilities.
We thank all NOIRLab and Blanco staff for their flexibility and support in helping to co-schedule all DECAT programs.

The DECAT nightly observing plan in 2021A and B semesters was primarily generated by co-authors Gautham Narayan, Dillon Brout, and Armin Rest.
The 2021A DECAT Observers were: Lauren Aldoroty, Segev BenZvi, Dillon Brout, Sasha Brownsberger, Melissa Graham, Chris Lidman, Gautham Narayan, Antonella Palmese, Armin Rest, Tom Shanks, Monika Soraisam, Kathy Vivas, Jiawen Yang, Qian Yang, Zhefu Yu.
The 2021B DECAT Observers were: Dillon Brout, Tony Chen, Alice Eltvedt, Shenming Fu, Melissa Graham, Chris Lidman, Gautham Narayan, Justin Pierel, Armin Rest, Ryan Ridden-Harper, Tom Shanks, Monika Soraisam, Elana Urbach, Qinan Wang, Zhefu Yu, Alfredo Zenteno.

The DDF data processing and alert production pipeline was assembled, implemented, and validated by co-author R. A. Knop.
It was modified from an earlier pipeline written by D. Goldstein \citep{2019ApJ...881L...7G}.
It makes use of \texttt{numpy} \citep{harris2020array} and \texttt{scipy} \citep{2020SciPy-NMeth}; \texttt{pandas} \citep{mckinney-proc-scipy-2010} for data management and statistical functions;  \texttt{astropy},\footnote{\url{https://www.astropy.org}} a community-developed core Python package for astronomy \citep{2018AJ....156..123T}; \texttt{astroquery} \citep{2019AJ....157...98G} and the datalab Query Client \citep{2014SPIE.9149E..1TF} for querying external astronomical databases, \texttt{SCAMP}  \citep{2006ASPC..351..112B} and \texttt{SWARP}  \citep{2002ASPC..281..228B} for catalog matching and image registration; \texttt{SExtractor} \citep{1996A&AS..117..393B} to extract sources from images; \texttt{HOTPANTS} \citep{2015ascl.soft04004B} for image subtraction; and \texttt{sqlalchemy} \citep{sqlalchemy} as a SQL database interface.

The ZADS alert distribution system was operated by co-authors E. Bellm and C. A. Phillips. 
We gratefully acknowledge support for ZADS from the Heising-Simons foundation under grant \#2018-0908 and thank Maria Patterson, Mario Juric, and Spencer Nelson for their contributions to the construction of ZADS.

This project used data obtained with the Dark Energy Camera (DECam),
which was constructed by the Dark Energy Survey (DES) collaboration.
Funding for the DES Projects has been provided by 
the U.S. Department of Energy, 
the U.S. National Science Foundation, 
the Ministry of Science and Education of Spain, 
the Science and Technology Facilities Council of the United Kingdom, 
the Higher Education Funding Council for England, 
the National Center for Supercomputing Applications at the University of Illinois at Urbana-Champaign, 
the Kavli Institute of Cosmological Physics at the University of Chicago, 
the Center for Cosmology and Astro-Particle Physics at the Ohio State University, 
the Mitchell Institute for Fundamental Physics and Astronomy at Texas A\&M University, 
Financiadora de Estudos e Projetos, Funda{\c c}{\~a}o Carlos Chagas Filho de Amparo {\`a} Pesquisa do Estado do Rio de Janeiro, 
Conselho Nacional de Desenvolvimento Cient{\'i}fico e Tecnol{\'o}gico and the Minist{\'e}rio da Ci{\^e}ncia, Tecnologia e Inovac{\~a}o, 
the Deutsche Forschungsgemeinschaft, 
and the Collaborating Institutions in the Dark Energy Survey. 
The Collaborating Institutions are 
Argonne National Laboratory, 
the University of California at Santa Cruz, 
the University of Cambridge, 
Centro de Investigaciones En{\'e}rgeticas, Medioambientales y Tecnol{\'o}gicas-Madrid, 
the University of Chicago, 
University College London, 
the DES-Brazil Consortium, 
the University of Edinburgh, 
the Eidgen{\"o}ssische Technische Hoch\-schule (ETH) Z{\"u}rich, 
Fermi National Accelerator Laboratory, 
the University of Illinois at Urbana-Champaign, 
the Institut de Ci{\`e}ncies de l'Espai (IEEC/CSIC), 
the Institut de F{\'i}sica d'Altes Energies, 
Lawrence Berkeley National Laboratory, 
the Ludwig-Maximilians Universit{\"a}t M{\"u}nchen and the associated Excellence Cluster Universe, 
the University of Michigan, 
NSF’s NOIRLab, 
the University of Nottingham, 
the Ohio State University, 
the OzDES Membership Consortium
the University of Pennsylvania, 
the University of Portsmouth, 
SLAC National Accelerator Laboratory, 
Stanford University, 
the University of Sussex, 
and Texas A\&M University.

Based on observations at Cerro Tololo Inter-American Observatory, NSF’s NOIRLab (NOIRLab Prop. IDs 2021A-0113 and 2021B-0149; PI: M. L. Graham), which is managed by the Association of Universities for Research in Astronomy (AURA) under a cooperative agreement with the National Science Foundation.

Co-author C.~E. Mart{\'i}nez-V{\'a}zquez is supported by the international Gemini Observatory, a program of NSF's NOIRLab, which is managed by the Association of Universities for Research in Astronomy (AURA) under a cooperative agreement with the National Science Foundation, on behalf of the Gemini partnership of Argentina, Brazil, Canada, Chile, the Republic of Korea, and the United States of America. 
Support for M.C., K.B.-V., and N.R.-S. is provided by the Ministry for the Economy, Development, and Tourism's Millennium Science Initiative through grant ICN12\textunderscore 12009, awarded to the Millennium Institute of Astrophysics (MAS), and by Proyecto Basal ACE210002 and FB210003.  

\medskip
Software: 
Astropy \citep{2018AJ....156..123T}, 
astroquery \citep{2019AJ....157...98G}, 
DoPHOT \citep{DOPHOT1, DOPHOT2}, 
HOTPANTS \citep{2015ascl.soft04004B}, 
NOIRLab Query Client \citep{2014SPIE.9149E..1TF}, 
numpy \citep{harris2020array}, 
pandas \citep{jeff_reback_2022_6702671}, 
photpipe \citep{Rest2005,Rest2014}, 
SCAMP \citep{2006ASPC..351..112B}, 
Scikit-learn \citep{scikit-learn}, 
SExtractor \citep{1996A&AS..117..393B}, 
sqlalchemy \citep{sqlalchemy}, 
SWARP \citep{2002ASPC..281..228B}.

\section*{Data Availability}

The proprietary period for the DECam images obtained as part of the DDF program has been waived.
The images from 2021A and 2021B, as presented in this work, are all available in the NOIRLab archive under proposal identifiers 2021A-0113 and 2021B-0149 (P.I. M.~L. Graham).
The images for 2022A and 2022B are available under proposal identifiers 2022A-724693 and 2022B-762878 (P.I. M.~L. Graham).

Photometry for the 4413 "probably-real" candidates (both the individual difference-image and nightly-epoch photometry) are available in the Version 2 release of the {\tt decam\_ddf\_tools} repository at \url{www.github.com/MelissaGraham/decam_ddf_tools/tree/v2.0}.


\bibliographystyle{mnras}
\bibliography{myrefs}


\bsp	
\label{lastpage}
\end{document}